\documentclass[preprint,12pt,5p,times,twocolumn]{elsarticle}

\usepackage{amssymb,color}
\usepackage{colortbl,array,multirow}
\usepackage{epstopdf}
\usepackage{ulem}

\journal{Physica D}

\begin{document}

\let\ol=\overline
\let\dd=\partial
\let\la=\langle
\let\ra=\rangle
\let\f=\frac
\let\ds=\displaystyle

\newcommand{\be}{\begin{equation}}
\newcommand{\ee}{\end{equation}}
\newcommand{\bc}{\begin{center}}
\newcommand{\ec}{\end{center}}
\newcommand{\bq}{\begin{quote}}
\newcommand{\eq}{\end{quote}}
\newcommand{\noi}{\noindent}			
\newcommand{\bnabla}{\mathbf{\nabla}}
\newcommand{\disp}{\displaystyle}
\newcommand{\bu}{\bf{u}}
\newcommand{\bg}{\bf g}
\newcommand{\bn}{\boldmath \nabla}
\newcommand{\dr}{\left(\frac{\Delta\rho}{\rho_s}\right)}
\newcommand{\nab}{\mbox{\boldmath $\nabla$} {}}
\newcommand{\ui}{\overline u_i}
\newcommand{\uj}{\overline u_j}
\newcommand{\xxi}{x_i}
\newcommand{\xxj}{x_j}
\newcommand{\odr}{\left(\frac{\Delta\overline{\rho}}{\rho_s}\right)}

\newcommand{\cms}{\nobreak\mbox{$\;$cm.s$^{-1}$}}
\newcommand{\cmss}{\nobreak\mbox{$\;$cm$^2$.s$^{-1}$}}

\begin{frontmatter}



\title{Understanding and modelling turbulent fluxes and
entrainment in a gravity current}


\author[label1]{P. Odier}
\author[label2,label3]{J. Chen}
\author[label2]{R.E. Ecke}

\address[label1]{Laboratoire de Physique,
\'Ecole Normale Sup\'erieure de Lyon,
46, all\'ee d'Italie 69364 Lyon Cedex 07, FRANCE}

\address[label2]{Condensed Matter \& Thermal Physics Group and Center for Nonlinear Studies, Los Alamos National Laboratory, Los Alamos, NM 87545, USA}

\address[label3]{School of Mechanical Engineering,
Purdue University,
585 Purdue Mall,
West Lafayette, IN 47907, USA}

\begin{abstract}
We present an experimental study of the mixing processes in a gravity current flowing on an inclined plane. The turbulent transport of momentum and density can be described in a very direct and compact form by a Prandtl mixing length
model: the turbulent vertical fluxes of momentum and density are found to scale quadratically with the
vertical mean gradients of velocity and density. The scaling coefficient, the square of the mixing length, is approximately constant over the mixing zone of the stratified shear layer. We show how, in different flow configurations, this length can be related to the shear length of the flow $(\varepsilon/\partial_z u^3)^{1/2}$. We also study the fluctuations of the momentum and density turbulent fluxes, showing how they relate to mixing and to the entrainment/detrainment balance. We suggest a quantitative measure of local entrainment and detrainment derived from observed conditional correlations of density flux and density or vertical velocity fluctuations.
\end{abstract}

\begin{keyword}
mixing \sep gravity current \sep mixing length \sep entrainment \sep oceanic circulation


\end{keyword}

\end{frontmatter}


\section{Introduction}
\label{sec:intro}

Mixing in stratified shear flows is an important process in many geophysical situations. {Ferrari and Wunsch have detailed in~\cite{Ferrari:ARFM:09} the kinetic energy budget occurring over the large range of oceanic scales. Several physical mechanisms, not all completely understood, allow the energy to be transferred from large scale geostrophic motion to the very small scales where irreversible mixing can take place.} Of 
particular current interest is the mixing and entrainment of oceanic overflows where gravity currents take place in particular regions of the oceans (Denmark Strait flow, Mediterranean outflow). Although they occur in very localized regions, these currents, through their mixing processes, contribute 
strongly to the transport of heat and salinity in the global ocean via the thermohaline ``conveyor 
belt" \cite{Willebrand:PO:01,Price:Science:93}. The mixing processes occur at scales too small to be captured by the numerical simulations of this circulation, requiring a sub-grid parametrization. In situ measurements \cite{Girton:JPO:03}, as well as experimental studies, are necessary to provide an accurate description. In order to obtain a valid parametrization from a laboratory experiment, there is also a need for a model that extrapolates the parametrization to oceanic conditions.

{Ellison and Turner~\cite{Ellison:JFM:59} were the first to investigate this phenomenon experimentally, in the case of a current flowing into a homogeneous ambient medium, without rotation. They derived a model for the bulk properties of the flow, based on measurements over various quantities. Decades later, Baines~\cite{Baines:JFM:01,Baines:JFM:05} reproduced the experiment, with an ambient stratified flow, while more recently, Cenedese~\cite{Cenedese:JFM:08} studied gravity current in a rotating frame. 

{In a broader context, several studies have been devoted to the evolution of turbulence in a stratified shear flow. In such cases, the flow was generally not buoyancy-driven. Some of them, either numerically~\cite{Jacobitz:JFM:97} or experimentally~\cite{Rohr:JFM:88,Piccirillo:JFM:97}, were focusing on general properties of turbulence, such as the influence of stratification and shear intensity on the growing or decaying character of turbulence. Others looked at mixing properties in a similar way as we did~\cite{Pardyjak:JFM:02,Strang:JFM:01,Strang:JPO:01}. We give elsewhere~\cite{Odier:JFM:14} a comparison of our results with some of these studies.} But in all these cases, for the experimental studies, the type of measurements used were either qualitative (video with dye) or quantitative but pointwise (conductivity probe or density measured locally on a fluid sample). And in general, no velocity measurement was performed. In order to better understand and model the processes involved in turbulent mixing induced by gravity currents, it is essential to measure velocity and density fields over a spatially extended region, as is now allowed by techniques such as particle image velocimetry and laser induced fluorescence.}

We developed an experimental apparatus where an experimental gravity current can be created with the ability to vary the flow conditions such as the degree of stratification or the initial turbulence level.  Among the key quantities for describing turbulent mixing processes are the fluxes of turbulent transport of momentum and density, which can be computed via the correlation terms between fluctuating components of the velocity field or between one component of velocity and density. One important feature of our experiment is the ability to measure simultaneously high resolution velocity and density fields, thus allowing us to obtain such correlations. In the next section, we present the experimental set-up. Then, in section \ref{sec:eddy}, we use our measurement of the turbulent fluxes to determine eddy diffusivities in our mixing processes. Section \ref{sec:mixing_length} recalls and extends the results of an earlier publication~\cite{Odier:PRL:09}, where we showed that a mixing length model provides a better description of how the turbulent fluxes are related to local mean gradients, compared to the constant eddy diffusivity hypothesis. In section \ref{sec:scaling}, we use data taken in different parameter configurations to determine a scaling law for the measured mixing lengths. Finally, in section \ref{sec:fluct}, we use the PDF's of the fluxes to better understand the location of entrainment and detrainment processes in the gravity current, before concluding in section \ref{sec:conclu}.

\section{Experimental set-up}
\label{sec:setup}

\begin{figure}[h]
{\begin{minipage}{.47\textwidth}
\centering
\includegraphics[width=8.5cm]{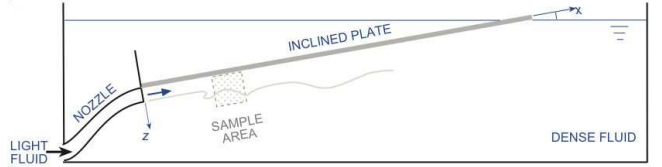}\\
\caption{\label{fig:device}  Sketch of the experimental device.}
\end{minipage}}
\end{figure}

The experiment, sketched in Fig.~\ref{fig:device}, and described in detail elsewhere \cite{Odier:JFM:14}, consists of a turbulent, uniform-density flow injected via a pump through a 5 cm high by 45 cm wide nozzle into a tank filled with unstirred higher density fluid.  The turbulence level of the injection current can be enhanced by an active grid device located just before the injection nozzle. The flow, upon exiting the nozzle, is bounded from above by a transparent plate inclined at an angle of 10$^{\rm o}$ with respect to horizontal, is unbounded below, and is confined in a tank about 2 m long, 0.5 m wide and 0.5 m high.  The components of the spatial position vector ${\bf x}$ describing the flow are the mean flow direction $x$, the cross-stream direction $y$ and the downward distance perpendicular to the plate $z$. The corresponding velocity ${\bf u(x)}$ has components $\lbrace u,v,w\rbrace$. We use the notation ${\bf \la u\ra}$ for a time- and ensemble-averaged quantity and ${\bf u'=u-\la u\ra}$ for its fluctuating part\footnote{In our previous publication, \cite{Odier:PRL:09},  $\overline{~\cdot~}$ represents ensemble averaging and $\langle~\cdot~\rangle$ denotes spatial averaging. This adjustment in nomenclature is made to be consistent with the popularly used ones in literatures, e.g. \cite{Pope:00}.}. The injection fluid, a solution of ethanol and water, is less dense than the fluid in the tank, water and salt (NaCl). This situation is reversed compared to oceanic overflows, where a denser fluid flows down an incline, but within the Boussinesq approximation, the physical mechanisms involved are the same. The density difference is defined as $\Delta\rho=\rho-\rho_{s}$, where $\rho$ is the measured density and $\rho_s$ is the maximum density, corresponding to the initial salt water density. The concentrations of ethanol and salt are adjusted (and the fluid temperatures maintained equal within 0.2$^\circ$C) so that the fluids are index matched to avoid optical distortions~\cite{McDougall:JFM:79}. {Each configuration presented in this study consisted in several runs with the same conditions, adding up to a total between 500 and 800 velocity fields.} All the fluids are freshly prepared for each run.

Instantaneous velocity and density fields are measured in a 9 cm $\times$ 9 cm area of a 0.1 cm thick laser sheet in the $x-z$ plane.  Velocity and density are measured simultaneously using particle image velocimetry (PIV) and planar laser-induced fluorescence (PLIF), respectively, at a rate of 3 Hz with two 2048$^2$ pixels digital cameras. Fluorescent dye (Rhodamine 6G) is added to the light fluid, and a calibration of density versus fluorescence intensity is performed for each position of the field of view. 

The lighter exit fluid is stably stratified with respect to the heavy fluid in the tank and forms a gravity current on the bottom side of the plate. The competition between the stabilizing effect of buoyancy and the destabilizing shear is captured in a dimensionless parameter, the Richardson number, $Ri = -(g\Delta\rho/\rho_s H)/(U^2)$ where $g$ is the acceleration of gravity, $U$ a typical velocity, and $H$ the height of the injection nozzle.  For small $Ri$, shear dominates buoyancy, and the flow is unstable to Kelvin-Helmholtz instability \cite{Thorpe:07}. In a standard configuration, used for the results in section \ref{sec:eddy}, \ref{sec:mixing_length} and \ref{sec:fluct}, the current is injected at a speed of $U_0$ = 7 cm/s, with the active grids on, and the initial density difference between the fluids is $\Delta \rho_0 = 2.6$ g/L. This results in an initial value of $Ri=0.27$ and a fully turbulent flow as it exits the nozzle with streamwise velocity fluctuations $u'$ about 25\% of $\la u\ra$, corresponding to a Taylor Reynolds number $Re_\lambda = u'^2/\sqrt{15\varepsilon\nu} \approx 100$, where $\nu$ is the fluid kinematic viscosity and $\varepsilon$ is the mean dissipation rate measured directly from velocity field (the spatial resolution of our velocity measurement is 0.5 mm compared to the dissipation scale of 0.33 mm).

Other configurations (see section \ref{sec:scaling}) were tested, where injection speed and/or initial density difference were varied, and in one case the active grids were removed. 

{In all configurations, there is rapid evolution of mean quantities over the first 20 cm. In this report, we focus on the region from 21 to 45 cm over which averages are approximately uniform along $x$. Note, however, that the results described here also apply to the initial region, except with a stronger dependence on downstream distance. A detailed description on the downstream evolution of various quantities associated to the current has been reported elsewhere~\cite{Odier:JFM:14}.}

\section{Eddy diffusivities}
\label{sec:eddy}

\begin{figure}[h]
{\begin{minipage}{.47\textwidth}
\centering
\includegraphics[width=7.8cm]{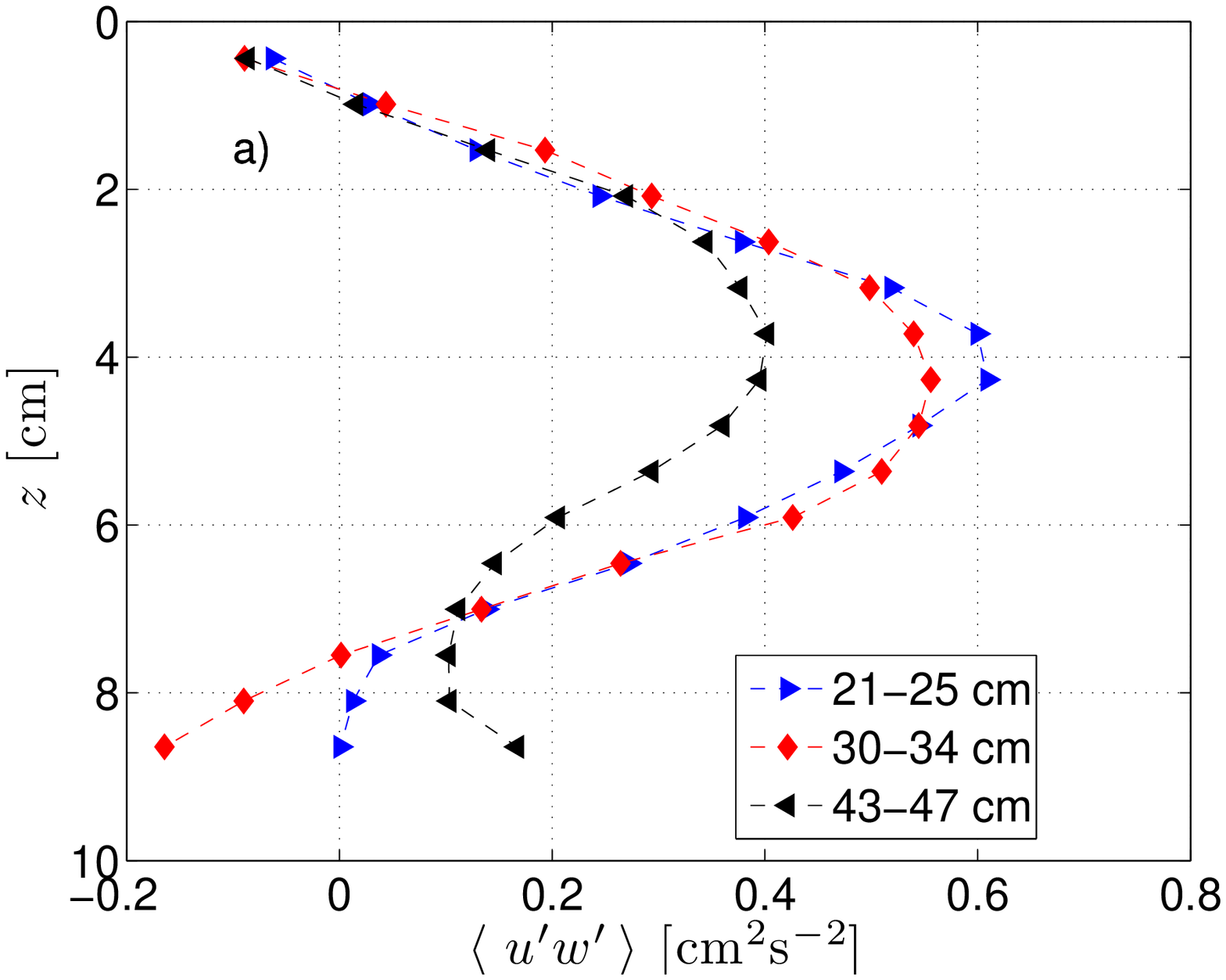}\\
\includegraphics[width=7.8cm]{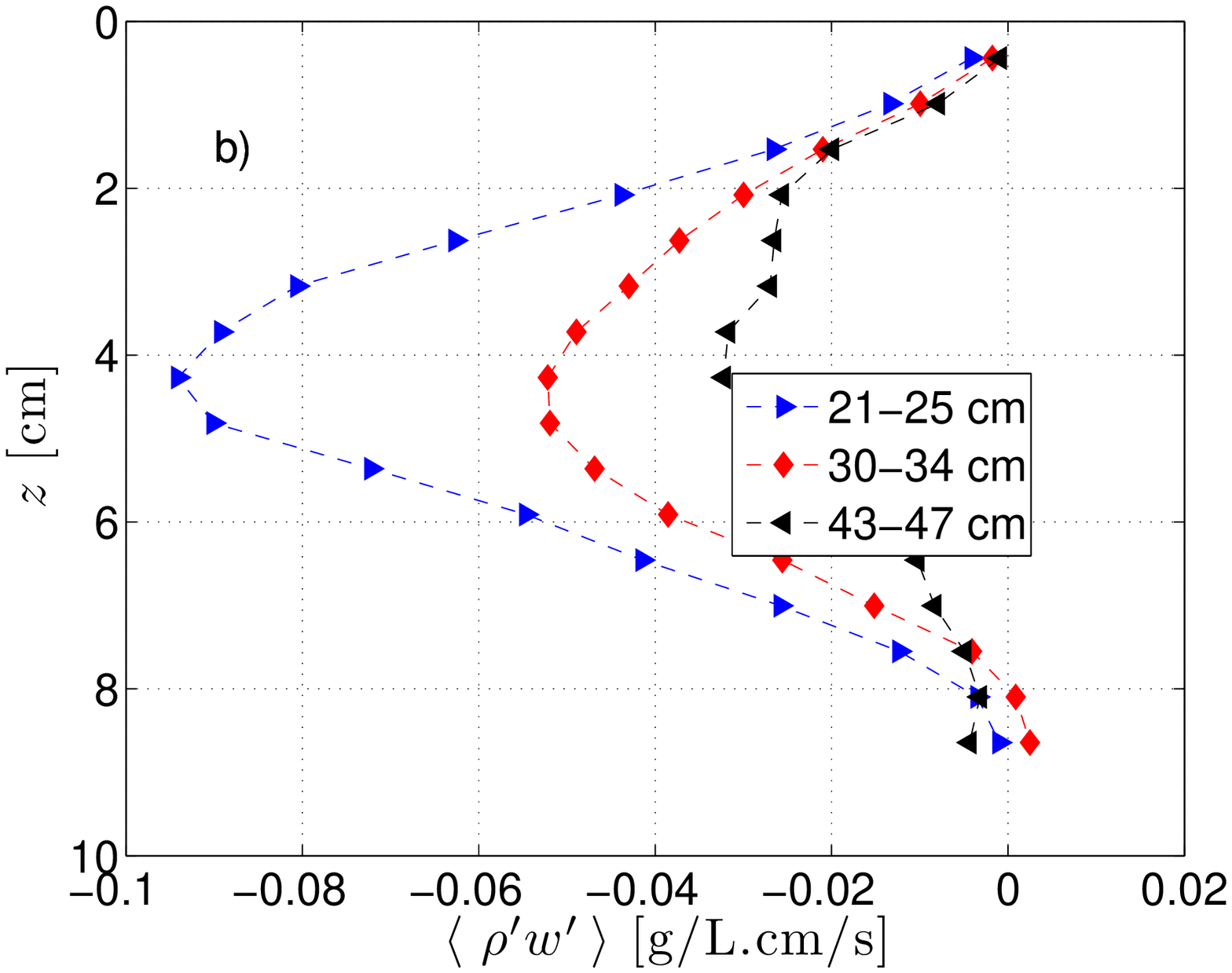}
\caption{\label{fig:flux}  Vertical profiles of the measured momentum turbulent flux (a) and density turbulent flux (b) measured at different distances downstream.}
\end{minipage}}
\end{figure}

As mentioned in the introduction, correlation products such as $\la{u'w'}\ra$ and $\la{\rho'w'}\ra$, which we can measure in our experiment, can be interpreted, respectively, as the vertical\footnote{Note that for simplicity we use the word ``vertical" for a direction that, strictly speaking, makes a 10$^{\rm o}$ angle with the vertical.} flux of downstream momentum and of density due to turbulent fluctuations. We show in Fig.~\ref{fig:flux} the vertical profiles of these two quantities computed locally as an ensemble average over all PIV/PLIF images and over all experimental runs. As expected, they display a maximum in the mixing region, close to the initial interface between the current and the ambient fluid. Close to the plate the mixing is reduced because perturbations that advect high concentration regions towards the plate are very rare, and far from the plate the ambient fluid is also undisturbed. In addition, the amplitude of the fluxes decreases as one goes away from the injection nozzle. This reduction reaches a factor of 1.5 for the momentum flux when the distance is doubled, and a factor of 3 for the density flux.

{In models of oceanic circulation, various closure schemes are used to parametrize turbulent transport (for a review, see \cite{Khanta:00,Burchard:02}). Some of these models are highly elaborate, closing the turbulence equations either at first moment level, providing sets of differential equations relating turbulent fluxes to the mean motion or even at second moment level, with equations for quantities like pressure strain-rate correlations. In more basic models (zero-equation models), the relation between turbulent fluxes and mean gradients is considered to be an algebraic one. For example, the model initially proposed by Pacanowski and Philander \cite{Pacanowski:JPO:81}, assumes a proportionality between fluxes and gradients, the proportionality constant being an effective diffusivity times a simple analytical function of the Richardson number. This proportionality constant is thus defined as $\nu_T=-\la{u'w'}\ra/\la\dd_z u\ra$ in the case of momentum transport and $\gamma_T=-\la{\rho'w'}\ra/\la\dd_z\rho\ra$ for density transport.} 

\begin{figure}[h]
{\begin{minipage}{.47\textwidth}
\centering
\includegraphics[width=8cm]{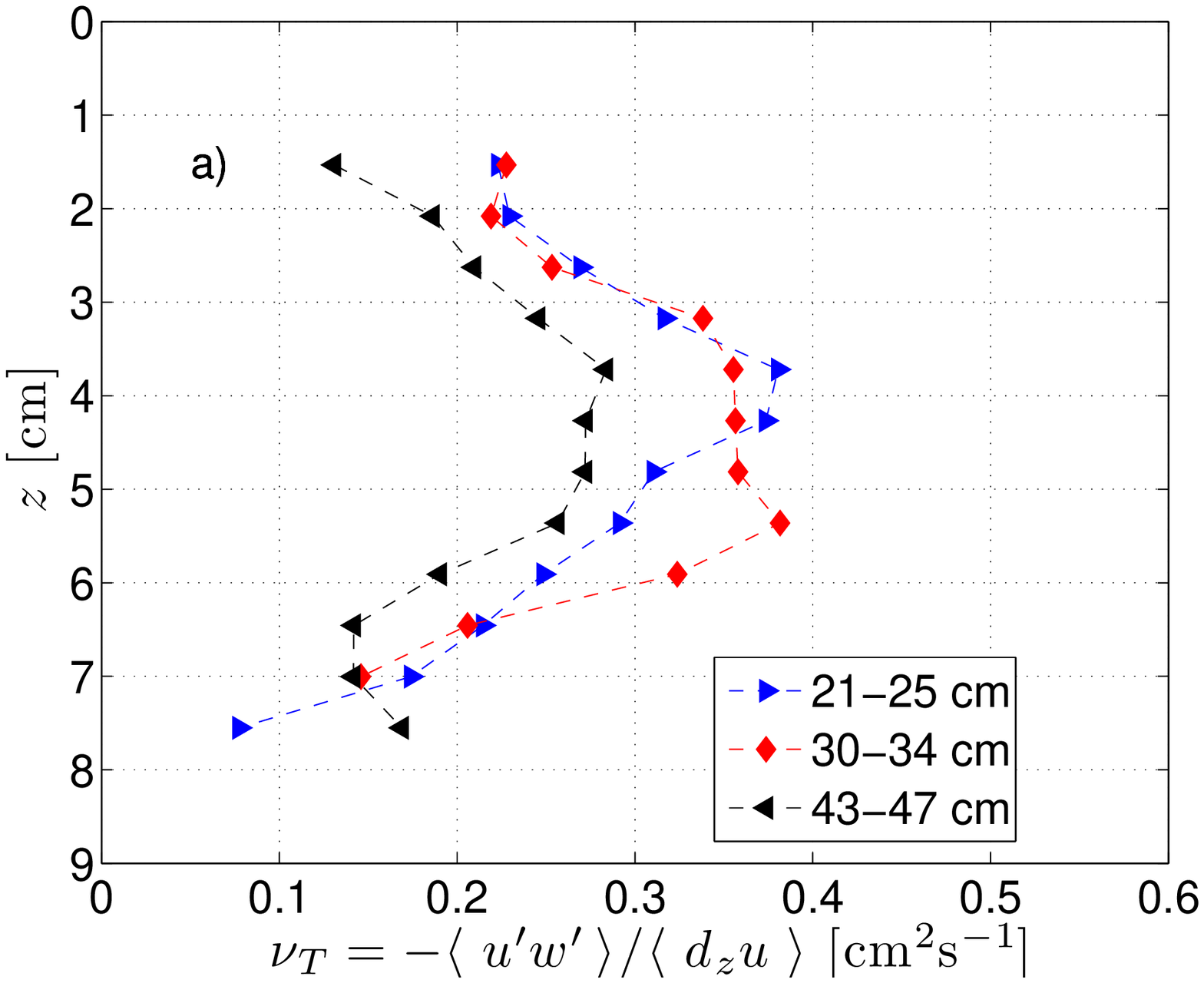}\\
\includegraphics[width=8cm]{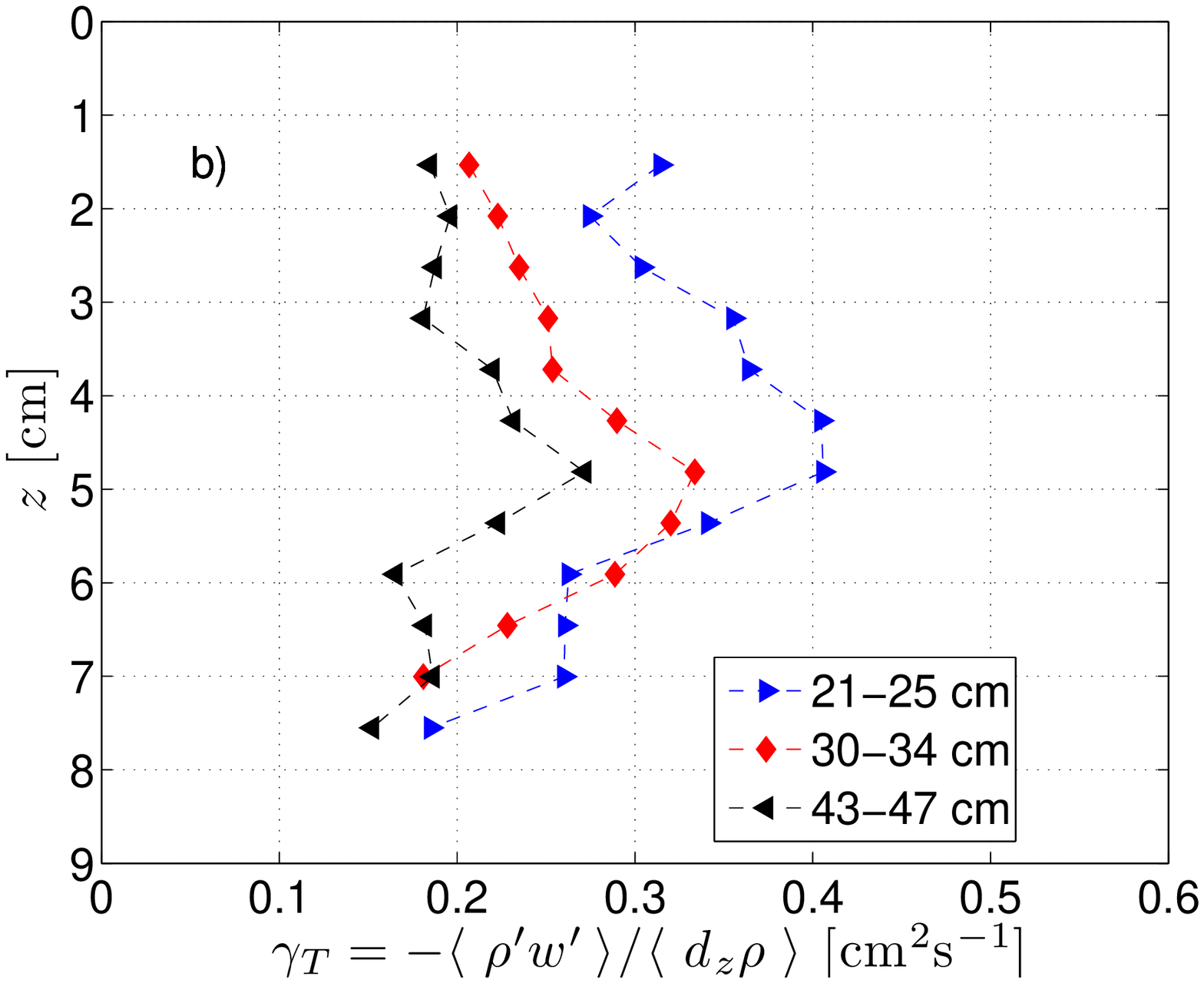}
\caption{\label{fig:eddy}  Vertical profiles of the measured eddy viscosity (a) and eddy density diffusivity (b) computed at different distances downstream.}
\end{minipage}}
\end{figure}

Our respective measurements of the fluxes and the corresponding gradients allow us to make an experimental determination of the eddy diffusion coefficients. Figure~\ref{fig:eddy} shows the vertical profiles of the measured eddy diffusivities, in the case of momentum transport (coefficient $\nu_T$) and density transport (coefficient $\gamma_T$). They are not shown for extreme locations, too close to the plate ($x<1.5$ cm) and too far away from the plate ($x>7.5$ cm), where the mixing is too weak to allow reliable measurement.They are displayed for 3 different distances from the injection nozzle.

It is interesting to note that the measured value of both coefficients is the same, about 30 times the molecular viscosity. In addition, these values vary by a factor of 3 between the central region, where most of the mixing occurs, and the top and bottom regions, which are respectively the less disturbed region in the current near the inclined plate and the less disturbed region in the ambient fluid away from the current. The coefficients also seem to decrease as one measures them further away from the injection nozzle. 

\begin{figure}[h]
{\begin{minipage}{.47\textwidth}
\centering
\includegraphics[width=8.5cm]{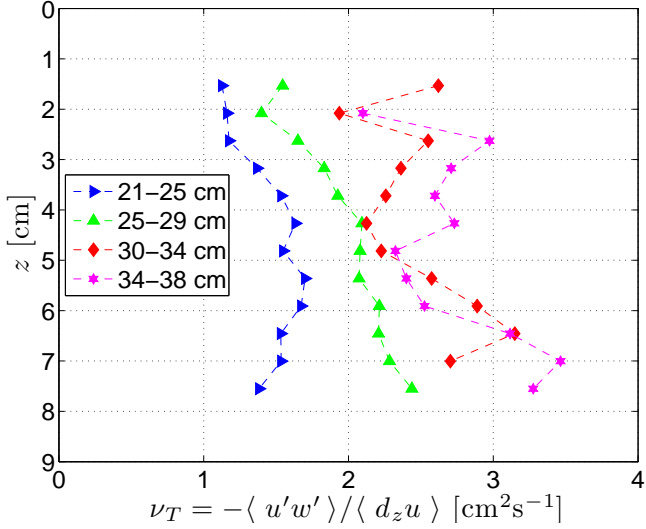}\\
\caption{\label{fig:eddy_NS}  {Vertical profiles of the measured eddy viscosity, computed at different distances downstream, for an unstratified flow. Because of the very different downstream evolution in this case, some measurements are shown for distances downstream different from the ones used in the other figures, and are therefore shown with different colors and symbols.}}\end{minipage}}
\end{figure}

{For comparison, we show in Fig.~\ref{fig:eddy_NS} the corresponding plot (analogous to Fig.~\ref{fig:eddy}a) in a case where there is no density difference between the injected fluid and the ambient fluid. This configuration corresponds to a plane jet along a wall. One can observe that the dependence of the eddy viscosity with the distance to the wall is much weaker: the vertical variation is of the order of 30\%, instead of the factor 3 observed in the stratified case. This observation is consistent with what has been measured in the case of unstratified jets (see for example figure 5.10 in~\cite{Pope:00}). In the same reference, the study of a plane self-similar jet shows that the eddy viscosity is expected to increase with downstream distance, which is what we observe in Fig.~\ref{fig:eddy_NS}, but not in Fig.~\ref{fig:eddy}. That the constant eddy viscosity model works better in the unstratified flow and fails rather badly in our stratified flow experiments is interesting but we are unable to explain this difference.}

\section{Mixing length model}
\label{sec:mixing_length}

The strong dependence of the turbulent diffusivities with depth and distance downstream calls for a better scheme to model the relation between fluxes and mean gradients. In an earlier publication \cite{Odier:PRL:09}, we showed that the observed scaling between these quantities is actually : $\la{u'w'}\ra\propto \la\partial_z  u\ra^2$ and $\la{\rho'w'}\ra\propto\vert  \la\partial_z  u\ra\vert\partial_z \rho$. We also demonstrated that this observation can be understood in the framework of Prandtl mixing length theory \cite{Prandtl}. Prandtl's argument is analogous to that applied in the kinetic theory of gases to molecular transport processes: it assumes that the coefficient of eddy viscosity is equal to the product of a ``mixing length" $L_m$, characteristic of the mixing phenomena, by a suitable velocity: $\nu_T  \simeq  L_m\times U({\rm typical})$. Assuming that $L_m$ is small enough so that the variation of the gradient over a distance $L_m$ can be neglected, one can take $U({\rm typical})=L_m\vert\la\dd_z u\ra\vert$ and thus obtain the relation $\la{u'w'}\ra=L_m^2\la\partial_z u\ra^2$, which is indeed observed (see figure 3 in~\cite{Odier:PRL:09}). The same argument for the density flux yields: $\la{\rho'w'}\ra=-L_{\rho}^2\vert\la\partial_z u\ra\vert\la\partial_z \rho\ra$, where $L_{\rho}$ is a mixing length associated with the density transport.

  As a result, we computed the mixing lengths as:


\be\label{equ:defLm}
L_m^2  =  \frac{\langle {u'w'}\rangle}{\langle \dd_z u\rangle^2}\;\;\; {\rm and~}
L_{\rho}^2  =  \frac{-\langle {\rho'w'}\rangle}{\vert\langle \dd_z u\ra\vert\la\dd_z\rho\rangle}
\ee

\begin{figure}[h]
{\begin{minipage}{.47\textwidth}
\centering
\includegraphics[width=8cm]{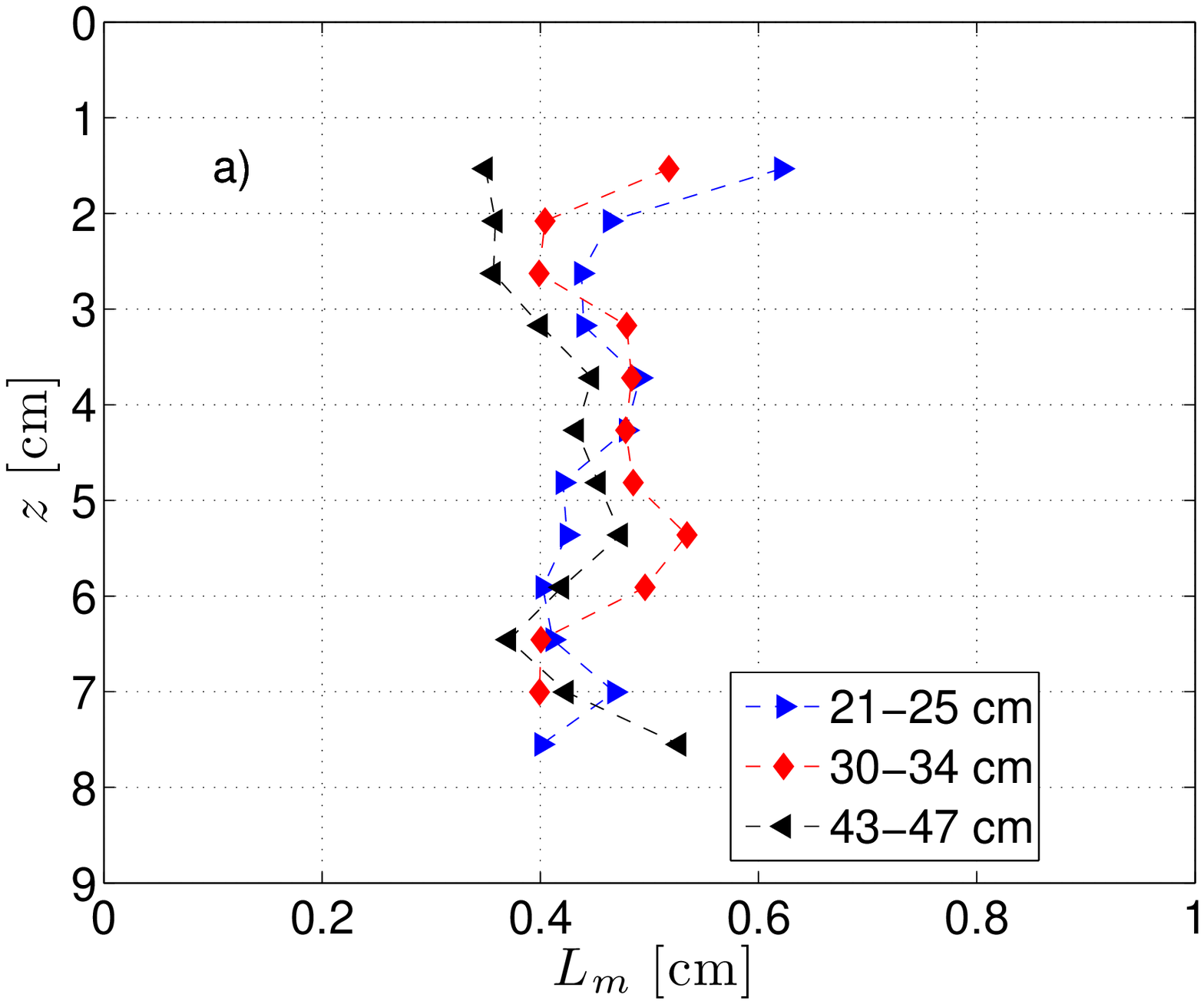}\\
\includegraphics[width=8cm]{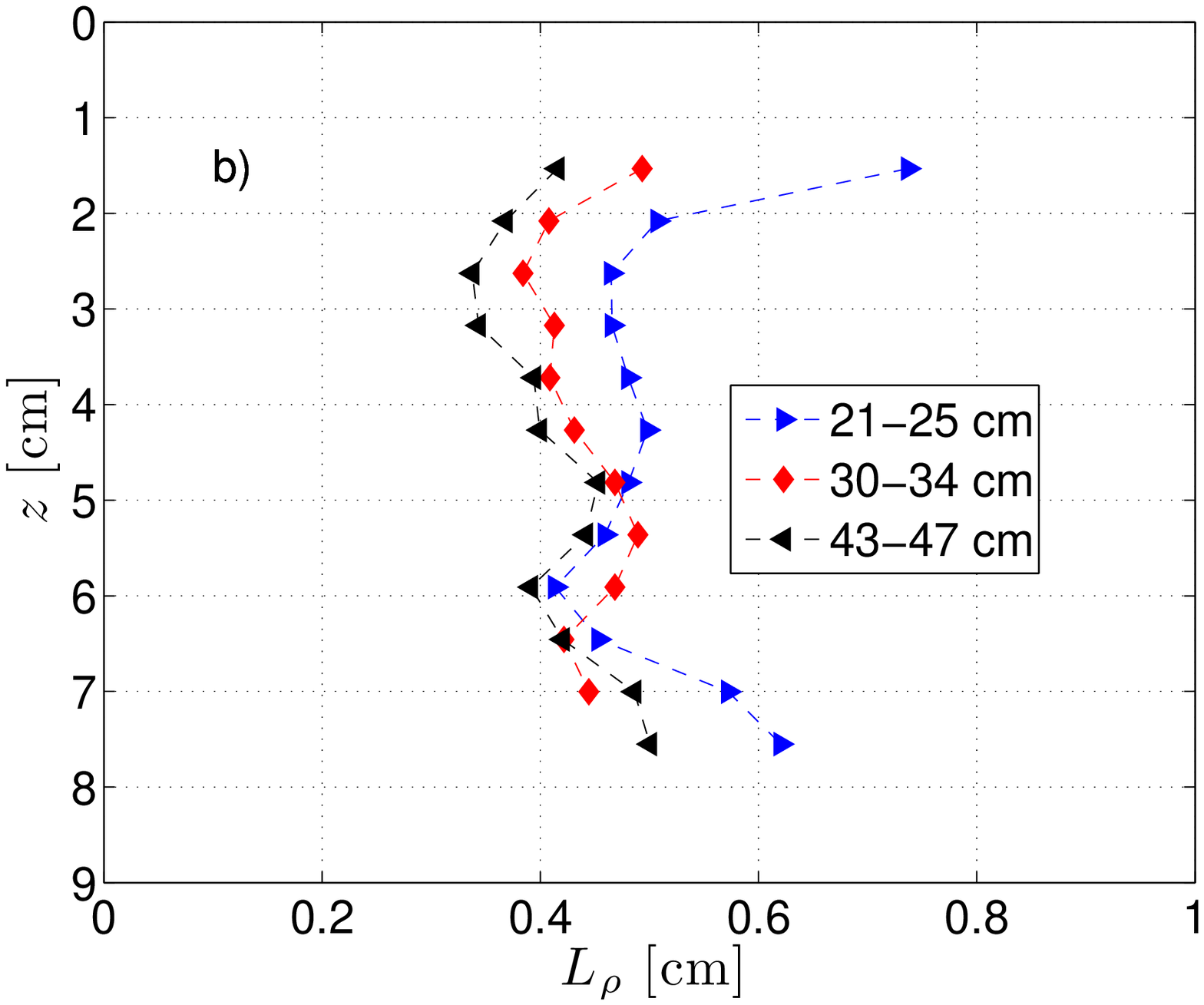}
\caption{\label{fig:mixing_lengths} Vertical profiles of the measured mixing lengths for momentum (a) and 
density (b) at different distances downstream. {Note that close to the plate, as well as far away from it, the gradients tend to vanish, producing large errors in the mixing length calculation. This explains the ``stray" data points which can be observed at the extremity of some curves.}}
\end{minipage}}
\end{figure}

The resulting vertical profiles of mixing lengths are shown in Fig.~\ref{fig:mixing_lengths}. They are much more uniform over depth, compared to the turbulent diffusivities profiles shown in Fig.~\ref{fig:eddy}. The dependence in downstream distance is also much weaker. {No decrease of the mixing lengths is observed as one approaches the wall ($z=0$), but one must remember that a law of the wall should only start within a distance from the wall of approximately 20\% of the flow width~\cite{Pope:00}, about 1 cm in our case, which is a region where we have no accurate measurement of the mixing lengths (see the remark about measurement errors in the caption of Fig.~\ref{fig:mixing_lengths}).} The mean values (average taken over depth and downstream distance) for both mixing lengths are the same: $L_m=L_{\rho}=0.45\pm 0.1$ cm

\section{Shear scale}
\label{sec:scaling}

In order to allow extrapolations of these results to oceanic conditions, it is necessary to understand how the mixing lengths depend on the degree of stratification of the current, as well as on its level of turbulence. In~\cite{Odier:PRL:09}, we derive an interpretation of the mixing length using the balance between the production of turbulent kinetic energy by the destabilizing shear and its dissipation. This balance defines a scale $L_s = (\ol\varepsilon/\ol{\la\dd_z u\ra}^3)^{1/2}$, which we call ``shear scale''~\cite{Smyth:PoF:00}. {This scale was defined initially by Corrsin~\cite{Corrsin:Naca:58} as the smallest scale at which anisotropy effects resulting from a large scale shear are carried out by the turbulent cascade. This scaling has been confirmed experimentally in a boundary layer flow~\cite{Saddoughi:JFM:94}.} As mentioned earlier, at low enough Richardson number, the effect of shear dominates the effect of buoyancy, therefore the relevant quantity to define the mixing scale is the shear and not the Brunt-V\"ais\"al\"a frequency $N = (g\ol{\la\partial_z\rho\ra}/\ol{\la\rho\ra})^{1/2}$. Since we are interested in a global scale for the flow, a spatial average ($\overline{~\cdot~}$ symbol) is taken for the quantities used to compute this scale, which corresponds to a different definition than the one used for the mixing lengths in equation \ref{equ:defLm}. This is important because the whole point of this section is to show that since the mixing lengths, although they are defined locally, do not depend on the local position, they can be related to a global quantity, namely the shear scale.

 We show in~\cite{Odier:PRL:09} that the value of the measured mixing lengths in the standard configuration is very close to the shear scale. In order to test the robustness of this observation, we measured the mixing lengths in various flow configurations. The results are displayed in table~\ref{tab:config}. Configuration 1 (shaded line in the table) is the standard configuration used until now (and also used for the data presented in \cite{Odier:PRL:09}). We then vary stratification and/or intensity of turbulence in 5 different configurations.  Since in all cases momentum and density mixing lengths are equal (except the non stratified case where $L_\rho$ is undefined), we give in column 7 the value of $L_m$ only.

\begin{table*}
\begin{center}
\begin{tabular}{|c|| l || l | l | l | l | l |l|l|l|}
\hline
\# & configuration & $\Delta\rho_0/\rho_s$ & $Ri$ & $U_0$ & $R_{\lambda}$ & $L_m$ 
& $\la{\ol \varepsilon}\ra$ & $\vert\la \ol{\dd_z u}\ra\vert$& $L_s$\\
~ & ~ & ~ & ~ & [cm/s] & ~ & [cm] 
& [cm$^2$/s$^3$] & [s$^{-1}$] & [cm]\\
\hline
\hline
\cellcolor[gray]{0.8} 1 & \cellcolor[gray]{0.8} standard & \cellcolor[gray]{0.8} 0.26\% & \cellcolor[gray]{0.8} 0.3 & \cellcolor[gray]{0.8} 7 & \cellcolor[gray]{0.8} 100 & \cellcolor[gray]{0.8} 0.45
&  \cellcolor[gray]{0.8} 0.8 &  \cellcolor[gray]{0.8} 1.4 &  \cellcolor[gray]{0.8} 0.55\\
\hline
2 & larger injection speed & 0.26\%& 0.2 & 9.5 & 140 & 0.6 & 1.4 & 1.4 & 0.7\\
\hline
3 & no active grid & 0.26\% & 0.35 & 6.5 & 42 & 0.35 & 0.7 & 1.7 & 0.35\\
\hline
4 & unstratified & 0 & 0 & 7.5 & 115 & 2.1 & 1.5 & 0.6 & 2.7\\ 
\hline
5 & double density & 0.52\% & 0.45 & 7 & 93 & 0.3 & 0.8 & 2.1 & 0.3\\
\hline
6 & double dens. & 0.52\% & 1.4 & 4.3 & 72 & 0.2 & 0.5 & 2.7 & 0.15\\
~ & half veloc. & ~ & ~ & ~ & ~ & ~ & ~ & ~ & ~\\
\hline
\end{tabular} 
\caption{Summary of the different experimental configurations. The first column is a reference number, the second gives the general features of the configuration, compared to the standard one (shaded line). The third column gives the initial density difference and the fourth the $Ri$. The fifth column shows the initial velocity of the current and the sixth shows $R_\lambda$. The 3 last columns give the data necessary to compute the shear length, and the shear length itself. The mean shear (column 9) is averaged over the half width of the vertical profile of the gradient.}\label{tab:config}
\end{center}
\end{table*}

Compared to case 1, cases 2 and 3 show that the value of the mixing length increases with turbulence intensity.\footnote{Note, however, that the quadratic relation between momentum flux and velocity gradient could not be clearly observed in the unstratified case because of a much larger spread of the data points, possibly owing to stronger mixing. This makes the corresponding calculation of the mixing length less reliable.} In the same way, as expected, cases 4 and 5 show that the stratification prevents mixing. In the last case, both the stratification is stronger and the shear weaker, resulting in a very short mixing length.

\begin{figure}[htb]
\centering
 \includegraphics[width=.48\textwidth]{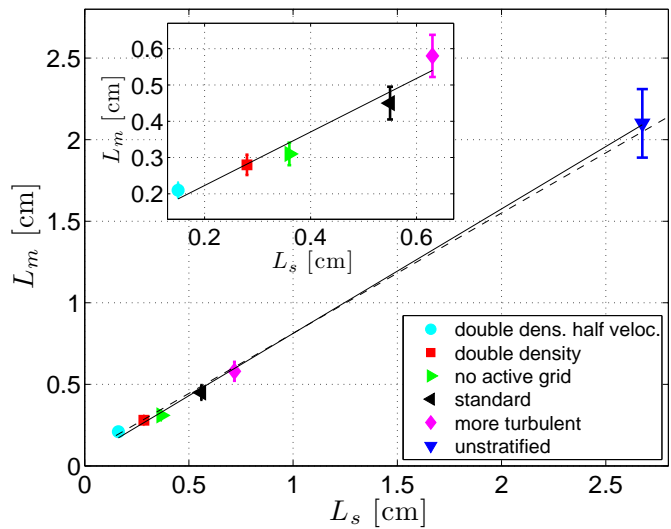}
\caption{Measured mixing length vs computed shear length, for the different flow configurations indicated. The error bars are determined by the standard deviation of the mixing length values at different locations in the flow. The solid line is a linear fit to the data. The inset shows a expanded view of the region below 1 cm, with a fit not including the point for the unstratified case. The slope for this fit is also shown in the main plot by a dashed line.}
\label{fig:scaling}
\end{figure}

  In Fig.~\ref{fig:scaling}, we plot the measured mixing length versus the computed shear scale, showing that in all cases studied, there is a scaling between the two quantities: $L_m=(0.7\pm 0.03) L_s$. In the unstratified case, we observed that the quadratic scaling between momentum flux and velocity gradient is less evident, {which may be related to the observation that the constant eddy viscosity model seems to work better in this case, as shown in section~\ref{sec:eddy}}. We included this data point in the graph however, since interestingly enough, it fits nicely the general trend, with a much larger value for the scales. One can see from the inset, which shows an expansion of small lengths (corresponding fit indicated in the main plot by a dashed line), that the slope is not very different (5\% change) when the unstratified point is included (or not) in the fit.
  
  {It may seem surprising that the mixing length scales with a quantity apparently independent of the stratification. But one must keep in mind that the mean turbulent dissipation rate $\ol\varepsilon$, and the mean shear $\ol{\la\dd_z u\ra}$ depend on the stratification. A study we performed, presented elsewhere~\cite{Odier:JFM:14}, shows that the shear is increased by the presence of stratification, because this stratification partially prevents the mixing that reduces the shear. This observation is in agreement with the observed decrease of the mixing length when the stratification is stronger. In the case of the turbulent dissipation rate, its dependence on the stratification is more subtle, since the buoyancy force acts at the same time as a turbulence suppressor and as a shear enhancer, since it is the driving force of the current.}

\section{Fluctuations of turbulent fluxes and entrainment}
\label{sec:fluct}

The mixing length model and the scaling properties presented in the previous sections allowed a better understanding of the average behavior of the turbulent momentum and density fluxes. In this section, we focus on the fluctuations of these fluxes, and how they relate to the mechanisms of entrainment and detrainment of the flow.  Entrainment occurs when a parcel of fluid that is heavier
(in our case) than the fluid within the gravity current and located outside that current is advected into the current and thoroughly mixed.  Average measures of entrainment show how much the gravity current increases in overall volume flow rate. Similarly, detrainment is the process in which a lighter element of fluid is transported by turbulent fluctuations into the heavier fluid outside the gravity current and is absorbed by mixing.  Below we present a novel method for characterizing the distribution of correlations of density and vertical velocity fluctuations that contribute to understanding the dynamics of the entrainment/detrainment process.

\begin{figure}
{\begin{minipage}{.47\textwidth}
\centering
\includegraphics[width=8cm]{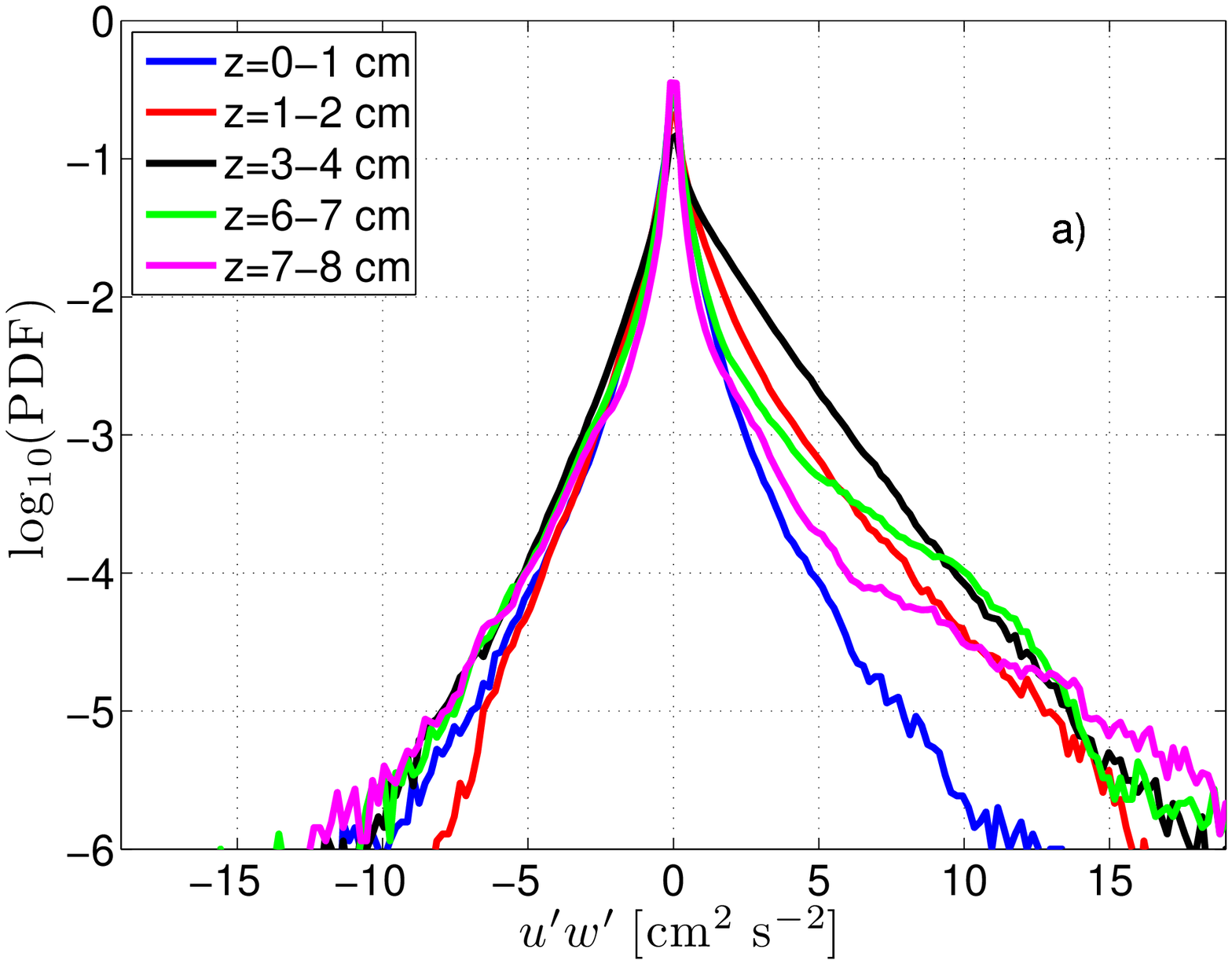}\\
\includegraphics[width=8cm]{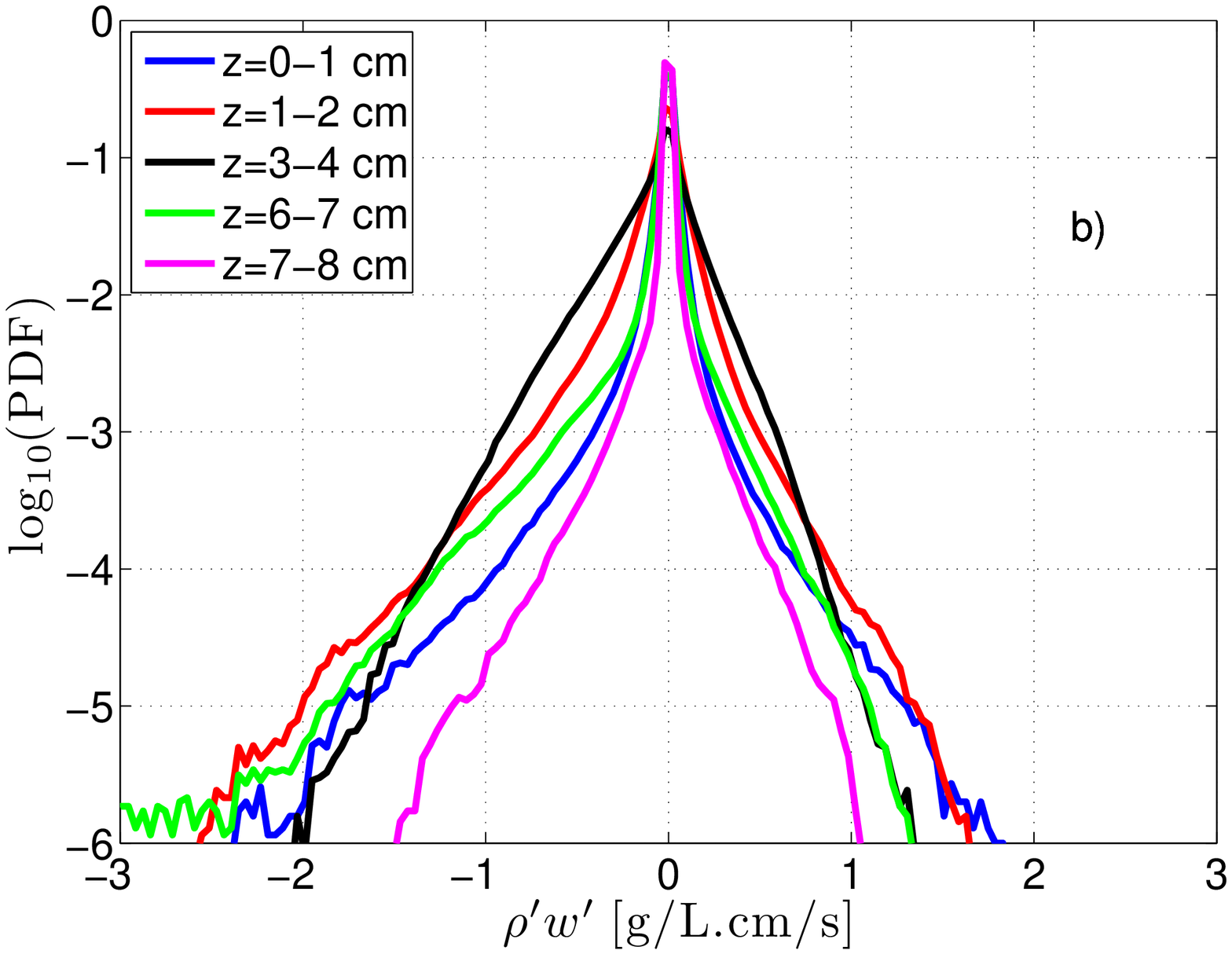}
\caption{\label{fig:hist} PDFs of the momentum flux (a) and density flux (b). Each PDF is constructed using data in a horizontal band of 1 cm height (vertical position indicated in the legend), situated between 20 and 49 cm from the injection nozzle.}
\end{minipage}}
\end{figure}

The probability density function (PDF) of the momentum and density fluxes, $\langle u' w' \rangle$ and $\langle \rho' w' \rangle$, respectively, are shown in Fig.~\ref{fig:hist}. We have to remind the reader that using our convention of signs for $w$ and $\rho$, positive $w'$ corresponds to a downward velocity fluctuation and positive $\rho'$ corresponds to a fluctuation of a particle heavier that the local mean. Each color corresponds to a given horizontal band, thus allowing one to see the evolution of the PDFs as $z$ increases. First, the fluxes reach large values compared to the mean as shown in Fig.~\ref{fig:flux}; there is still a probability $10^{-3}$ that a fluctuation will reach a value about 10 times the mean. As expected, this tendency to produce large fluctuations is stronger as one approaches the mixing region around the initial interface between the current and the ambient fluid (black curve, $z$=3-4 cm).

The PDFs of momentum and density fluxes are asymmetric. This asymmetry is the origin of the non-zero mean value of the fluxes. With our conventions of sign, it is positive for momentum and negative for density. In order to understand this asymmetry, it is necessary to give some considerations to the signs of the fluxes. As can be seen in Fig.~\ref{fig:device}, the $x$ axis is oriented in the direction of the flow and the $z$ axis points downwards. Thus, a positive value of $u'w'$ corresponds to either downward transport of downstream momentum or upward transport of upstream momentum. It is therefore understandable that the PDF of $u'w'$ displays more positive events, since these correspond to the standard transport of momentum owing to Kelvin-Helmholtz mixing. In other words, downstream fluctuations with downward transport take turbulent momentum from where it is large, namely in the interior of the mixing zone, towards the quiescent regions in the heavier fluid below the gravity current, i.e., at large $z$. In the same manner, negative values of $\rho'w'$ correspond either to upward transport of heavier fluid or to downward transport of lighter fluid. Either of these transport mechanisms oppose buoyancy and are caused again by Kelvin-Helmholtz mixing. This explains the negative asymmetry observed in Fig.~\ref{fig:hist}b. In addition, we observe that this asymmetry becomes stronger for the PDF in the center region (red and green, then black curve), where most of the mixing takes place. Finally, the PDFs of both momentum and density are broadest for the black curves, i.e., in the middle of the mixing zone.  On the other hand, the most extreme events occur on either side of the peak of the mixing zone where the average fluxes are largest.

\begin{figure}
{\begin{minipage}{.47\textwidth}
\centering
\includegraphics[width=8cm]{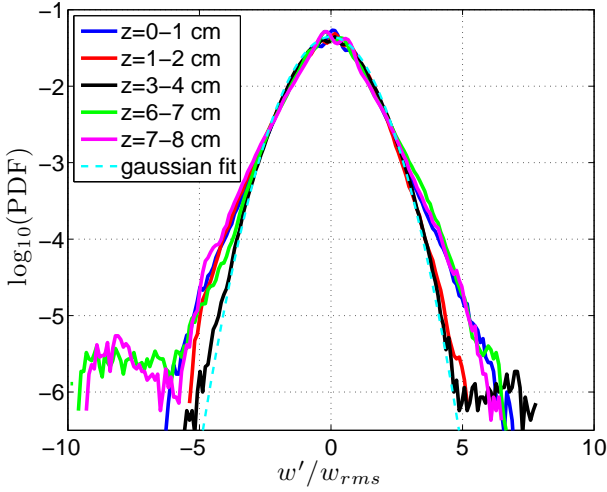}\\
\caption{\label{fig:Whist} Normalized PDFs of the vertical velocity fluctuations. Each PDF is constructed using data in a horizontal band of 1 cm height (vertical position indicated in the legend), situated between 20 and 49 cm from the injection nozzle.}
\end{minipage}}
\end{figure}

In order to compare the momentum and density flux PDFs to more standard probability distributions studied in turbulence, we show in Fig.~\ref{fig:Whist} PDFs of the vertical velocity fluctuations, normalized by their rms value, for horizontal layers at different distances from the plate. The PDFs are fairly Gaussian in the mixing region (black line, compared to a gaussian fit, indicated by a dashed line), whereas close to the plate or far away from it, the distributions differ from a Gaussian shape, with broader tails, probably owing to the intermittent and not fully-developed turbulence in these regions.

\begin{figure}
\centering
\includegraphics[width=8cm]{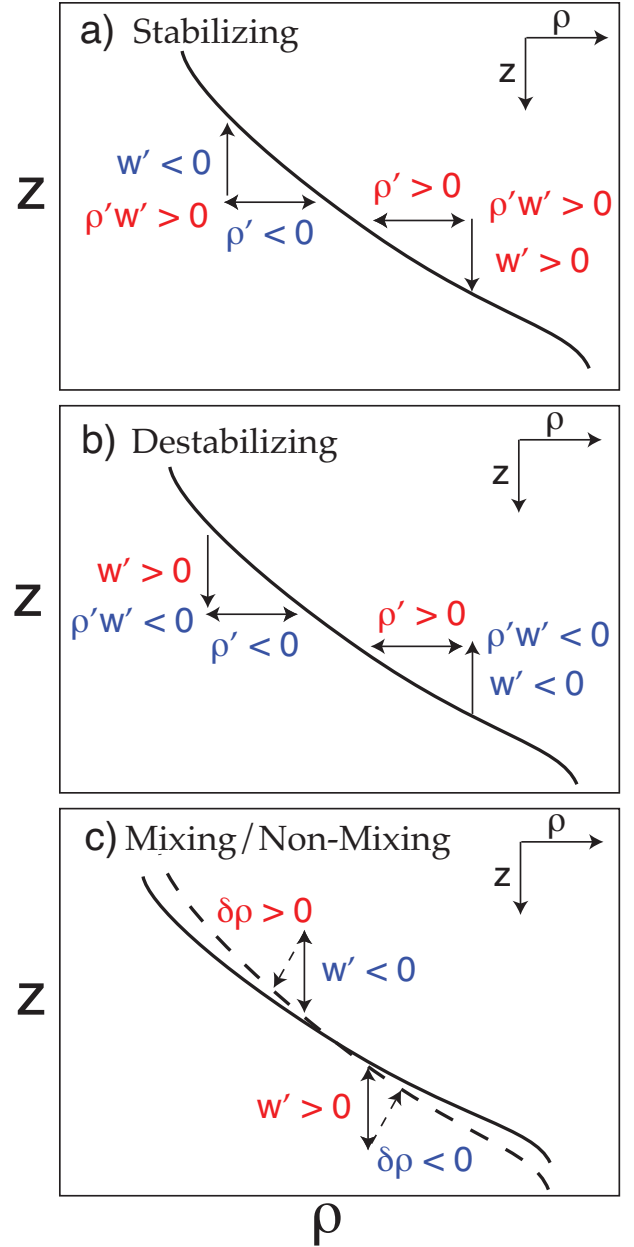}
\caption{\label{fig:quadrants} Schematic illustration of a) stabilizing velocity and density correlations resulting in $\rho' w' > 0$, b) destabilizing velocity and density correlations with $\rho' w' > 0$, and c) the net effect of mixing on fluid parcels that mix during a destabilizing displacement before gravitational forces can restore the parcel to its equilibrium vertical position of neutral buoyancy. In c), the solid (dashed) arrows in c) indicate non-mixing (mixing) processes, and the solid (dashed) curves show the non-mixing (mixing) mean density profile.}
\end{figure}

One can further elucidate the nature of the density flux by considering the instantaneous correlations between density fluctuations $\rho'$ and density flux $\rho'w'$. To understand these correlations, one needs to consider the expected behavior of a parcel of fluid in a background stratification defined by the average density $\bar{\rho}(z)$.  In the absence of turbulence, a parcel of fluid with positive or negative $\rho'$ should move towards its neutral buoyancy position, exchanging its potential energy for kinetic energy.  In the case we consider, a parcel of fluid that is lighter than its surroundings has a negative value of $\rho'$ and should move towards the plate, that is, have a negative vertical velocity $w' < 0$ so that the resultant density flux is positive $\rho' w' > 0$, whereas a parcel of heavier fluid has a positive $\rho'$ and would be expected to have a positive vertical velocity so again one has positive density flux, $\rho' w' > 0$. This situation is illustrated schematically in Fig.~\ref{fig:quadrants}a and indicates that for either positive or negative $\rho'$, stabilizing return to neutral
buoyancy corresponds to positive density flux. Here, and in further discussion below, we use an intuitive description based on tracking a fluid parcel in time, i.e., a Lagrangian perspective.  Our measurements, however, are Eulerian so that the parcel of fluid we consider at time $t$ is advected away by the mean flow.  In addition, the buoyancy condition implied by specifying a value of $\rho'$ is
not the instantaneous buoyancy experienced by a fluid parcel because $\rho'$ is determined by a long-time density average rather than
by the instantaneous distribution of density.  Nevertheless, the description presented here gives a reasonable understanding of our results and provides the basis for further exploration of these important correlations.

If we add turbulent fluctuations, the situation is a bit more complex, see Fig.~\ref{fig:quadrants}b.  A positive vertical velocity fluctuation $w' > 0$ may advect a lighter parcel of fluid with $\rho' < 0$ into a region of higher density, thereby producing a negative density flux $\rho' w' < 0$ and a condition which is unstable with respect to the potential energy of that parcel. Similarly, a negative value of $w' < 0$ advecting a parcel of fluid with $\rho' > 0$ also produces an unstable condition with $\rho' w' < 0$. Thus, destabilizing fluctuations have negative density flux $\rho' w' < 0$. 

\begin{figure*}
{\begin{minipage}{.95\textwidth}
\centering
\includegraphics[width=7.8cm]{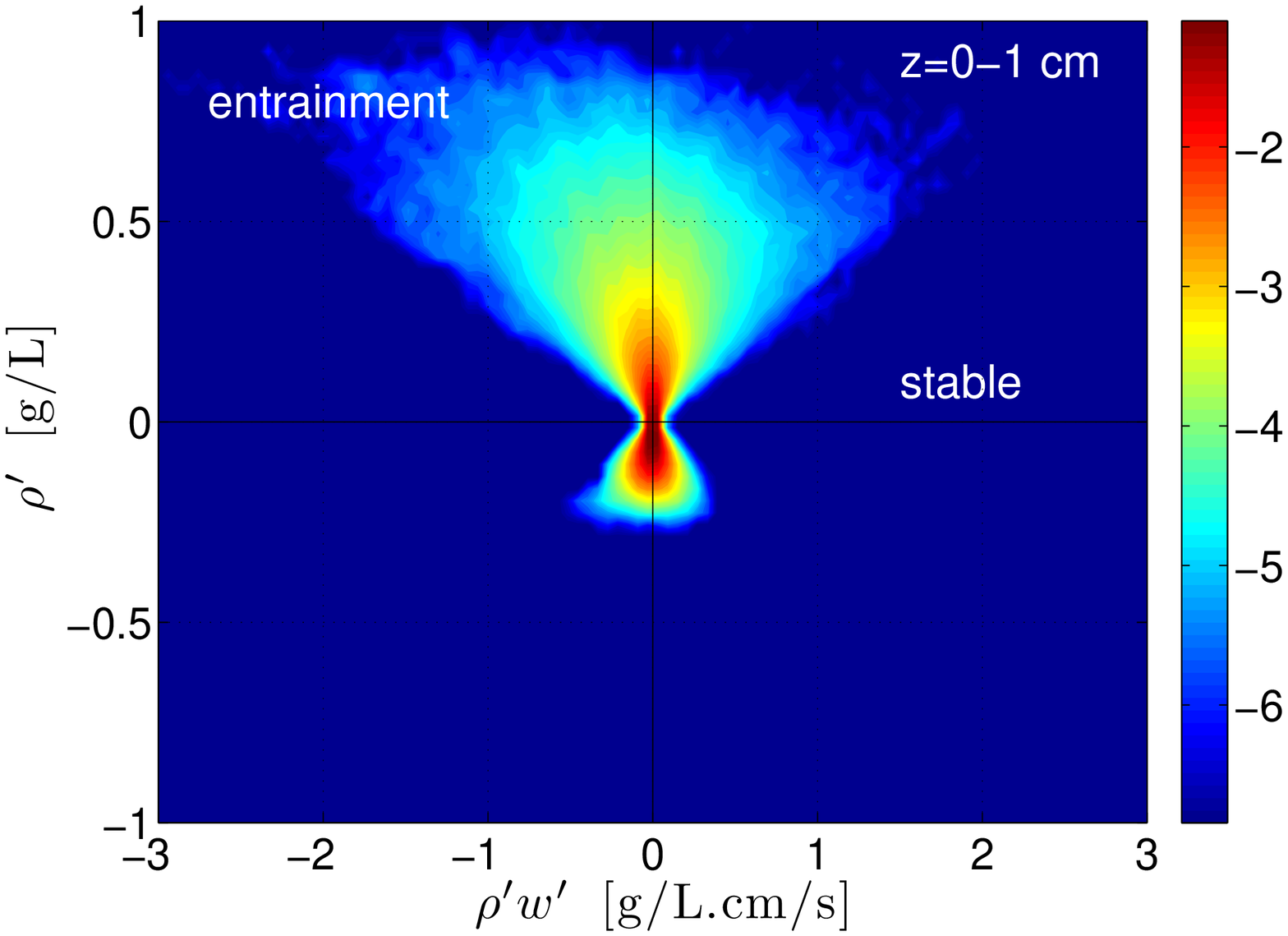}
\includegraphics[width=7.8cm]{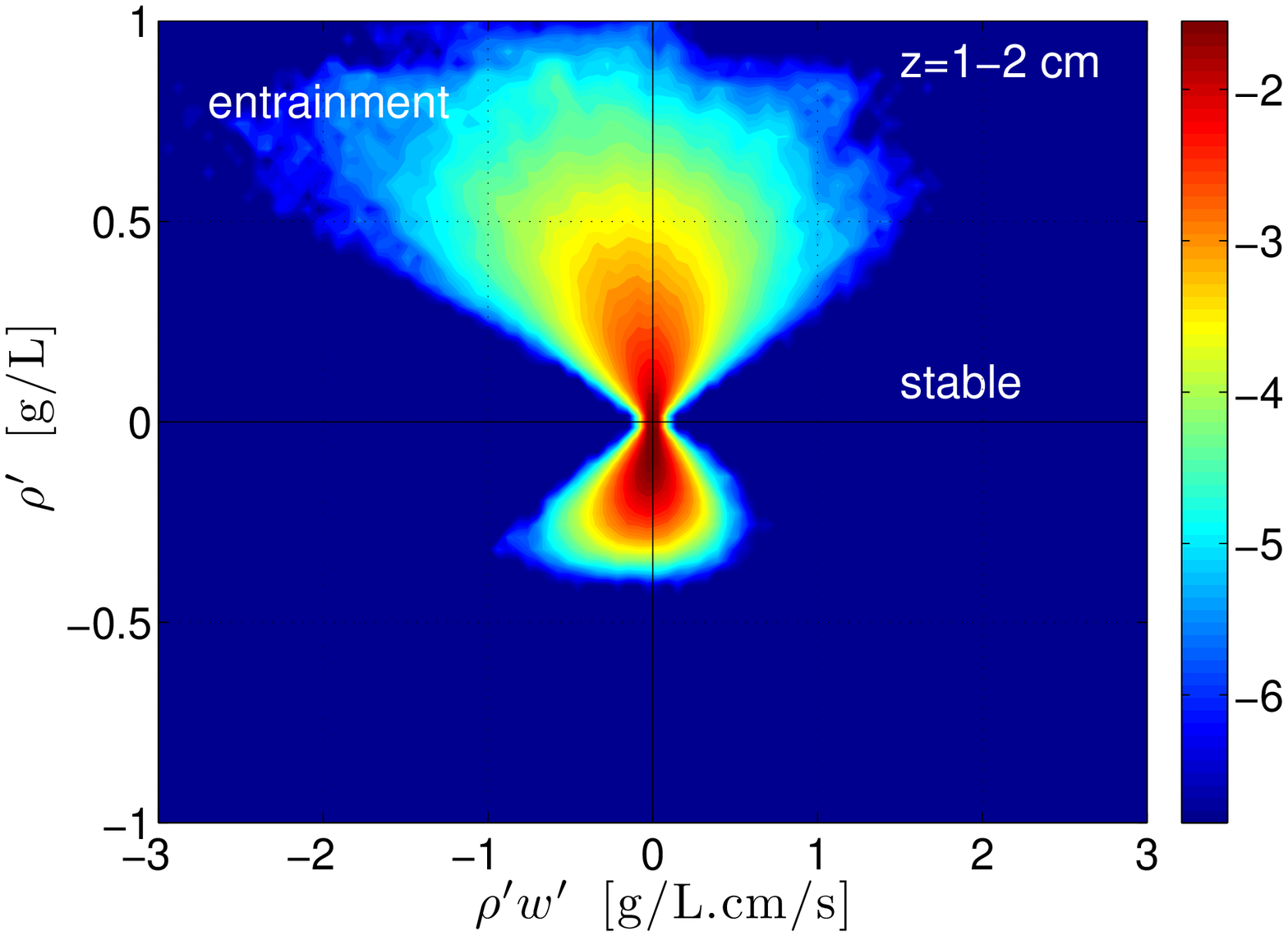}\\
\includegraphics[width=7.8cm]{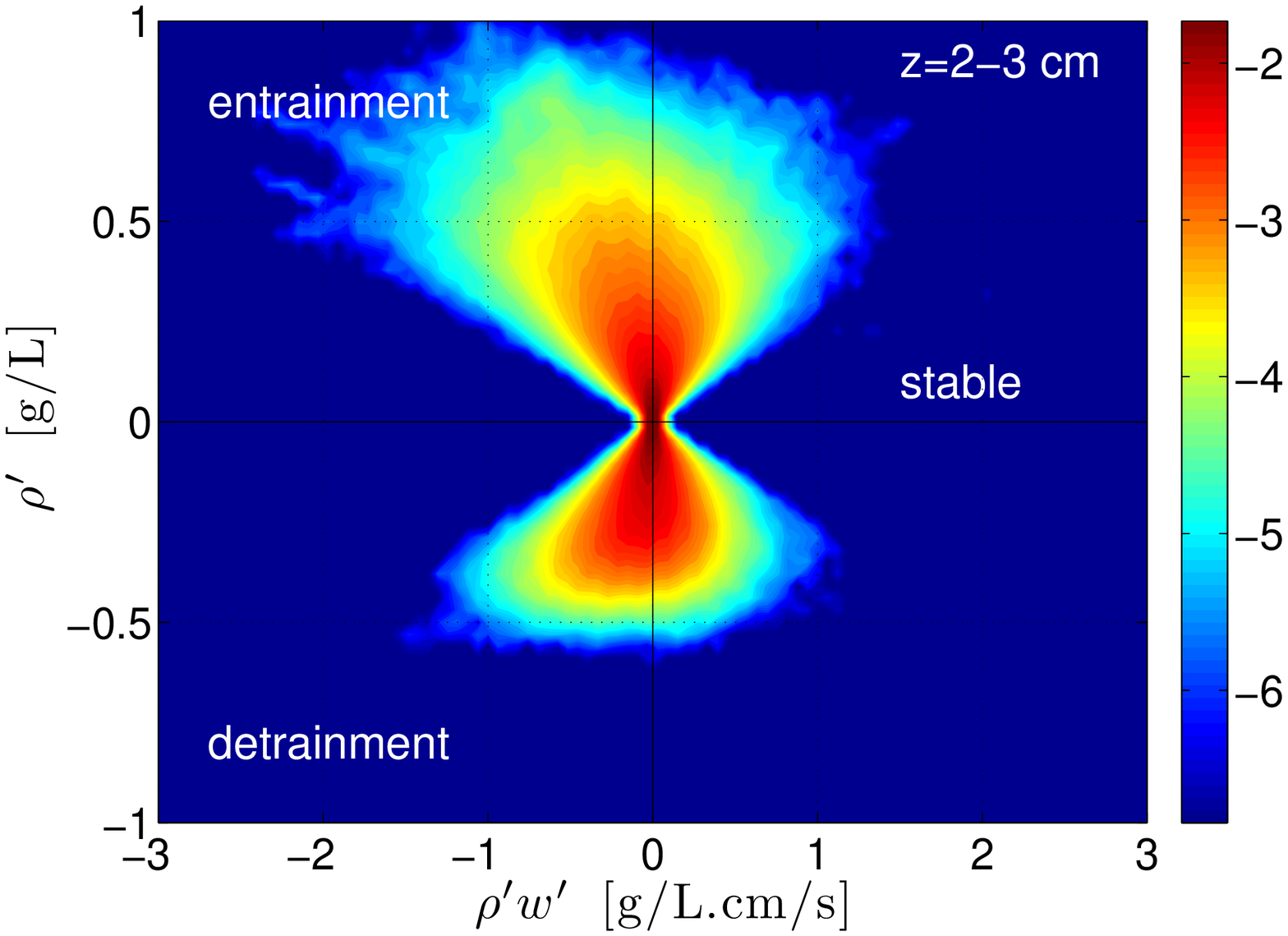}
\includegraphics[width=7.8cm]{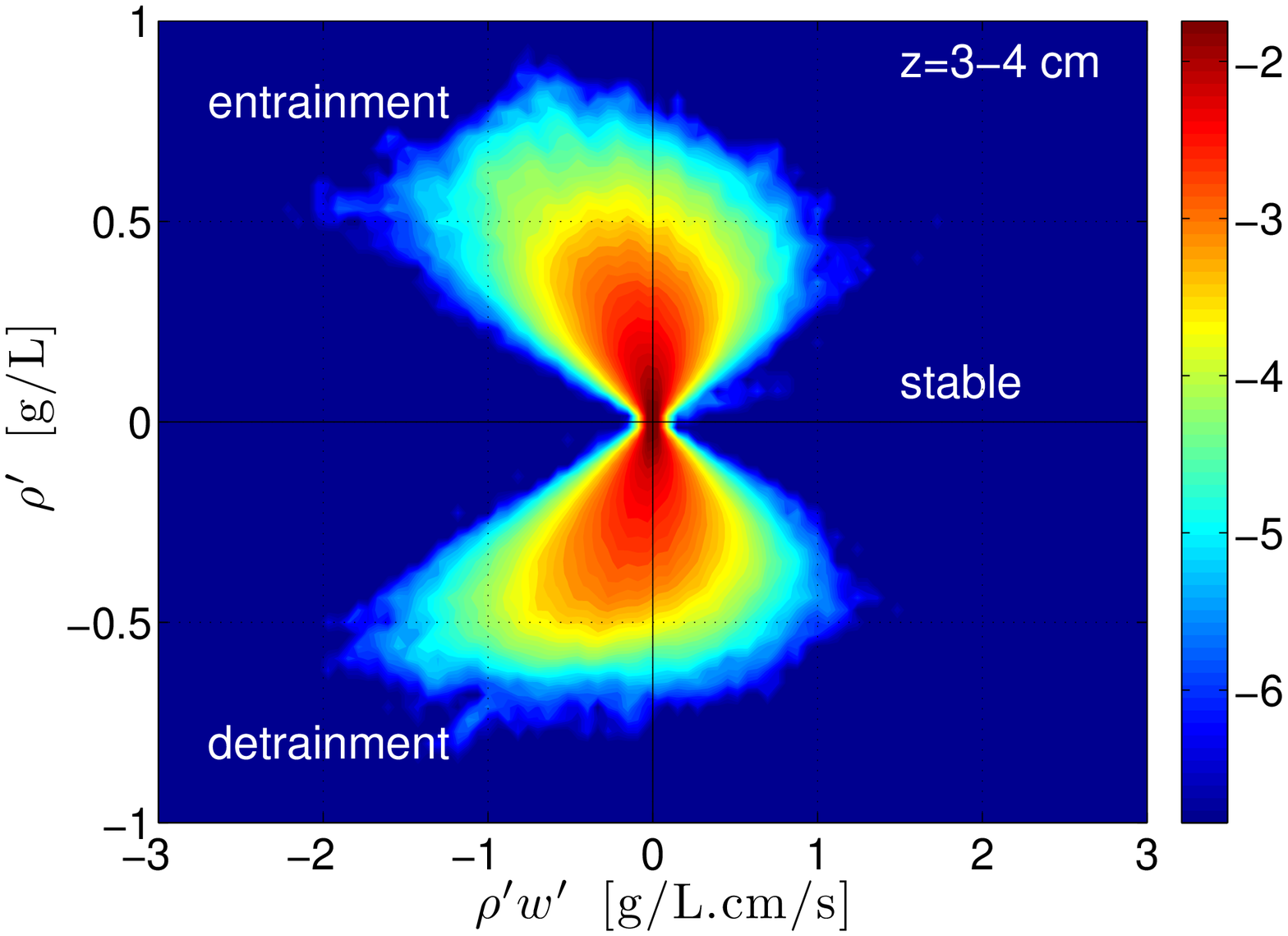}\\
\includegraphics[width=7.8cm]{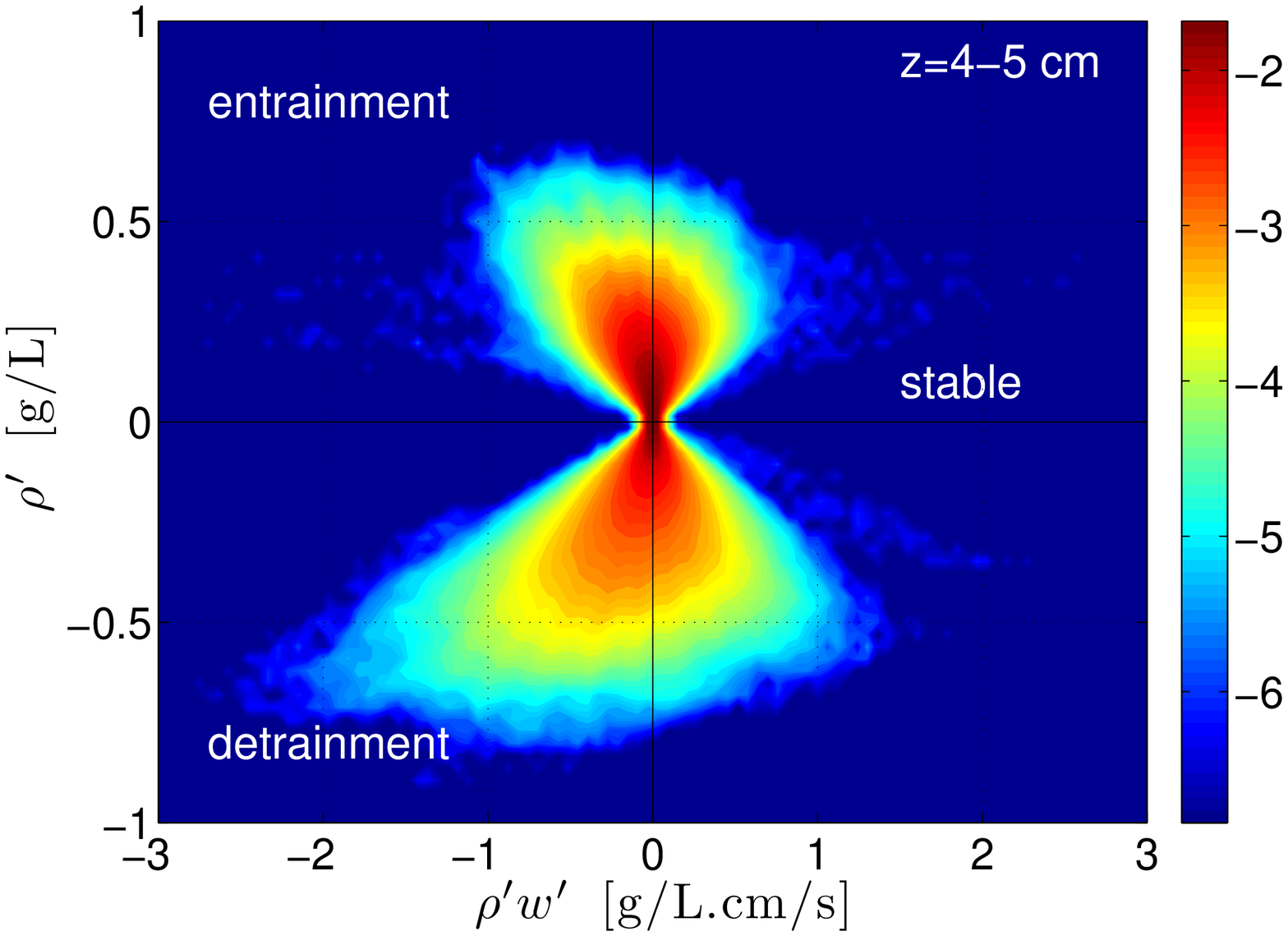}
\includegraphics[width=7.8cm]{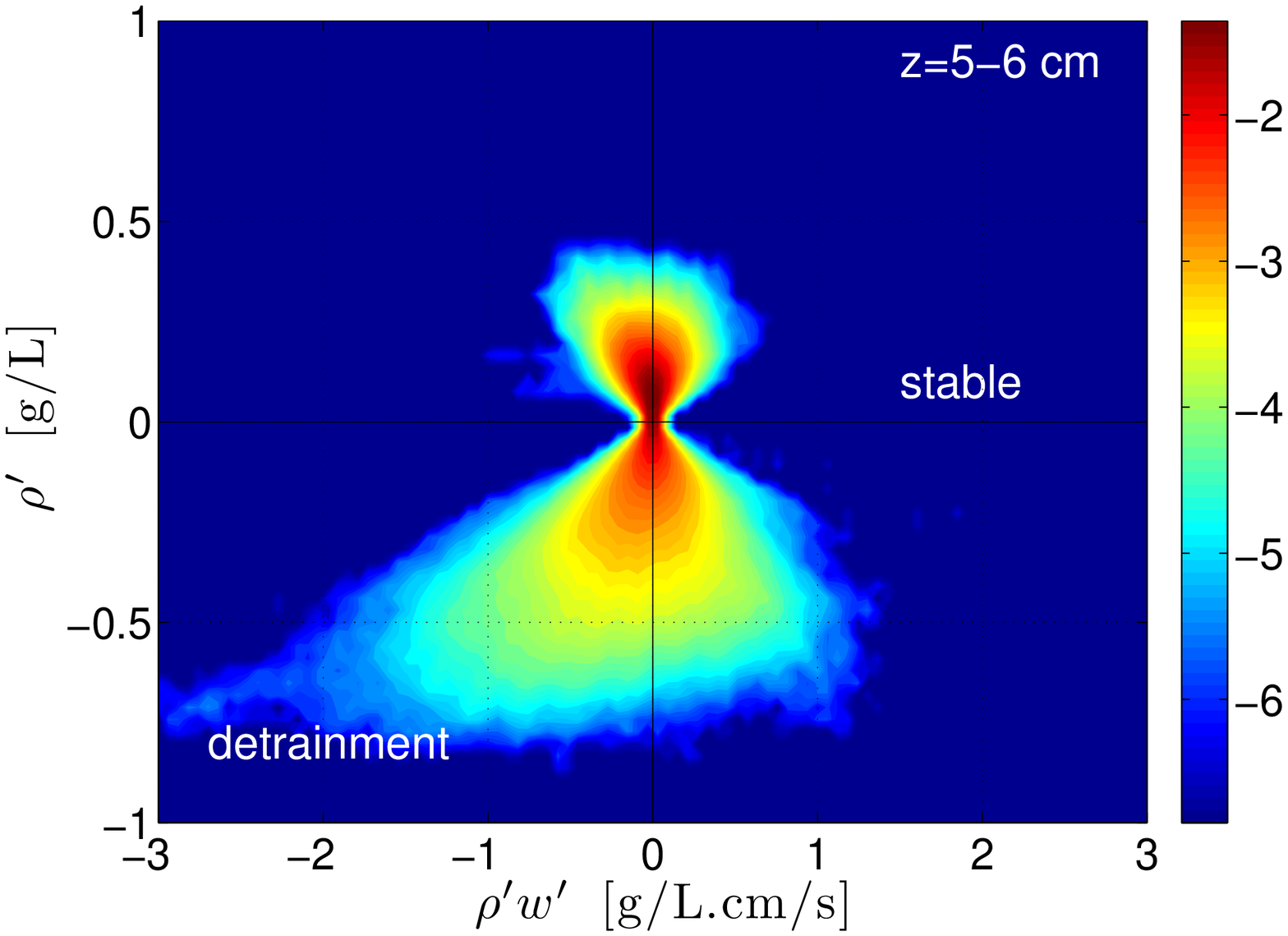}\\
\includegraphics[width=7.8cm]{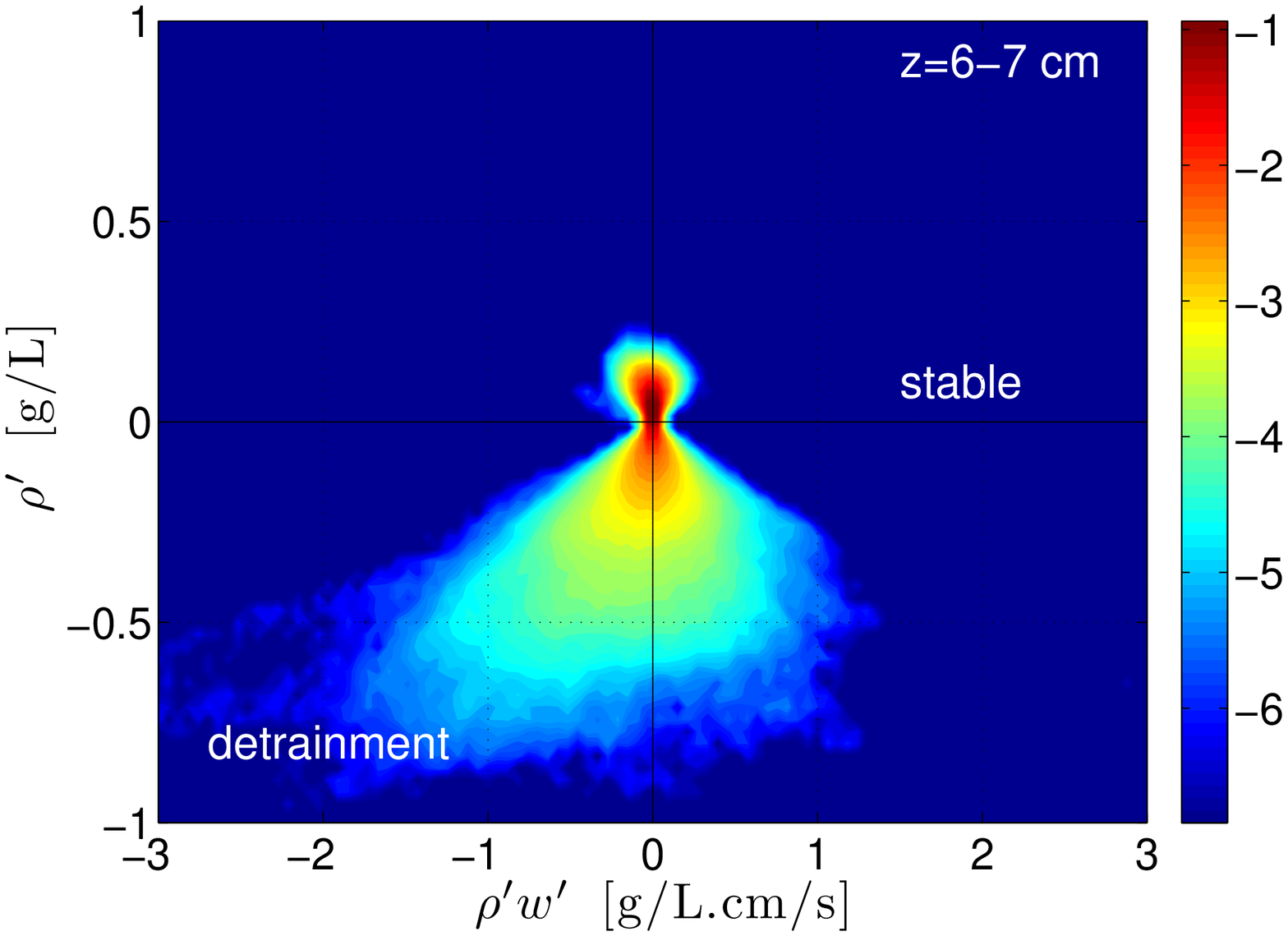}
\includegraphics[width=7.8cm]{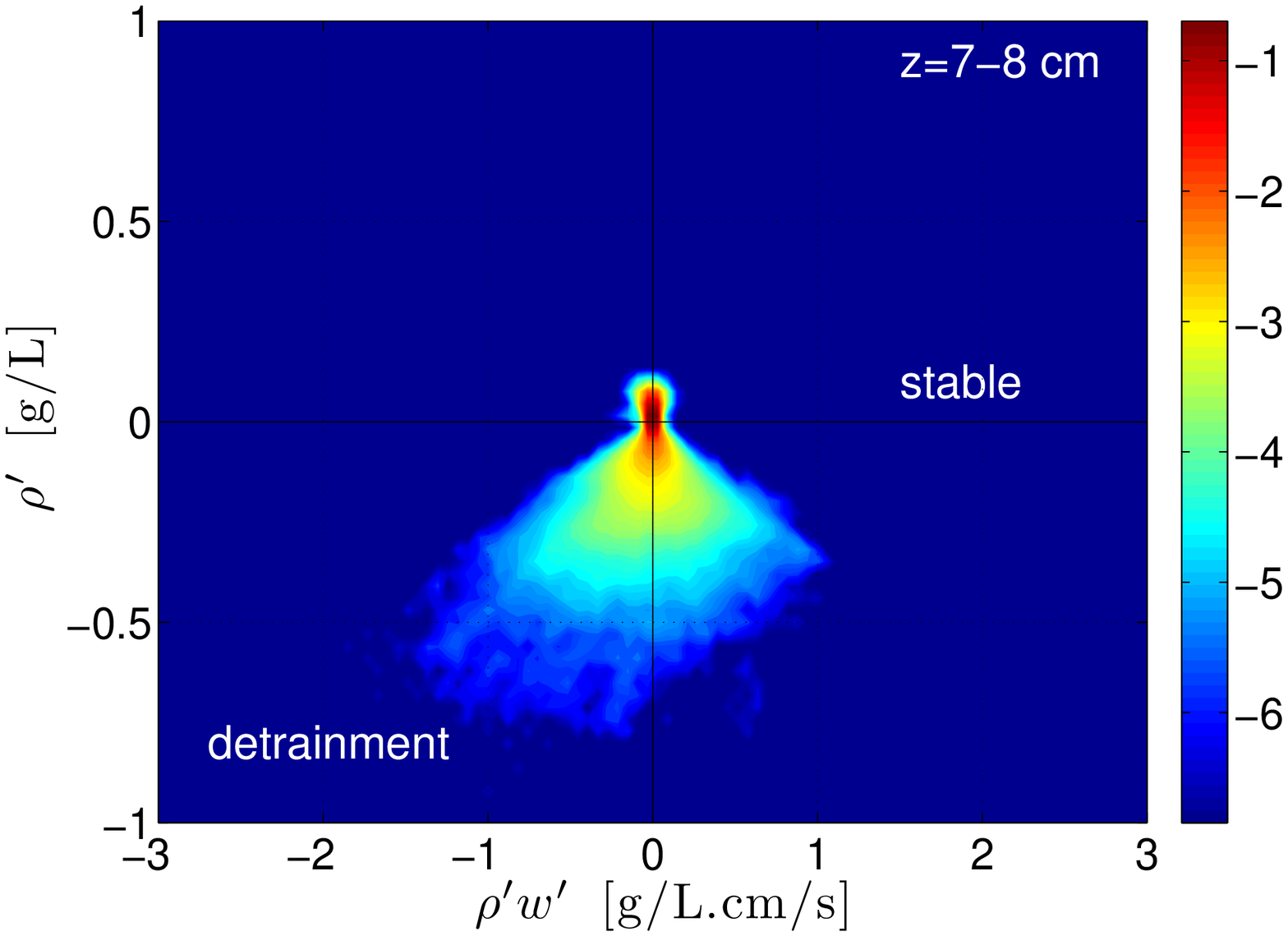}\\
\caption{\label{fig:2Dhist}  2D PDFs of the density flux vs density fluctuations. Each PDF is constructed using data in a horizontal band of 1 cm height (vertical position indicated in the top right corner of each plot), situated between 20 and 49 cm from the injection nozzle.}
\end{minipage}}
\end{figure*}

If no mixing occurs, these two processes of destabilizing and stabilizing motions should balance in that a parcel that is displaced vertically by turbulent fluctuations would be restored to neutral buoyancy owing to its adverse potential energy with respect to the mean density profile. This process is illustrated in Fig.~\ref{fig:quadrants}c by the solid arrows. If mixing occurs along this path, however, there will be a net effect on the local mean density profile that reduces the local density gradient as indicated in Fig.~\ref{fig:quadrants}c by the dashed arrows; a positive vertical velocity (downward) produces a net reduction in density upon mixing, i.e., $\delta \rho < 0$ whereas an initially negative $w$ will produce a net increase in density,  $\delta \rho > 0$. The overall effect of many such parcels being displaced away from neutral buoyancy, mixing partially or completely, and returning to a new neutral buoyancy vertical position is to reduce the global mean density gradient.  These examples described in  Fig.~\ref{fig:quadrants}c allow a reasonable interpretation of the experimentally obtained correlation PDFs between density fluctuations and density flux shown in Fig.~\ref{fig:2Dhist}. In particular, it allows one to obtain the amount of entrainment (fluid added to the gravity current) or detrainment (fluid left behind in the ambient fluid) of the flow as a function of vertical height. The balance between entrainment and detrainment in a boundary flow is an unresolved issue in numerical simulations \cite{Killworth:JPO:99} and may be resolved using the correlations between $\rho'$ and $\rho' w'$.

Fig.~\ref{fig:2Dhist} shows 2D PDFs of $\rho'$ versus $\rho'w'$, in each 1 cm horizontal band taken at different distances from the inclined plate. From the earlier discussion illustrated in Fig.~\ref{fig:quadrants}, we can tell that the right part of the plots in Fig.~\ref{fig:2Dhist} corresponds to the stabilizing return to neutral buoyancy (positive flux, see also Fig.~\ref{fig:quadrants}a), while the top left quadrant corresponds to entrainment (see right part of Fig.~\ref{fig:quadrants}b) and the bottom left quadrant to detrainment (see left part of Fig.~\ref{fig:quadrants}b). One observes that in the regions close to the plate, the distributions are fairly symmetric with respect to the sign of the fluxes, with only a slight negative skewness in the positive $\rho'$ lobe, indicating a small level of entrainment. Moving away from the plate, the asymmetry in the upper entrainment lobe increases, i.e., more entrainment, and the detrainment lobe with $\rho' < 0$ begins to become asymmetric as well. Near the middle of the mixing zone with $z \sim 3-4$ cm, the entrainment and detrainment lobes are about the same, indicating an approximately
equal amount of entrainment and detrainment.  Finally, far from the plate entrainment is inactive as indicated by the small and symmetric entrainment lobe compared to active detrainment demonstrated by the large and negatively skewed detrainment lobe.  The qualitative picture one gets from this analysis is that deep in the gravity current entrainment dominates whereas outside the current detrainment is dominant.

\begin{figure}
\centering
\includegraphics[width=8.2cm]{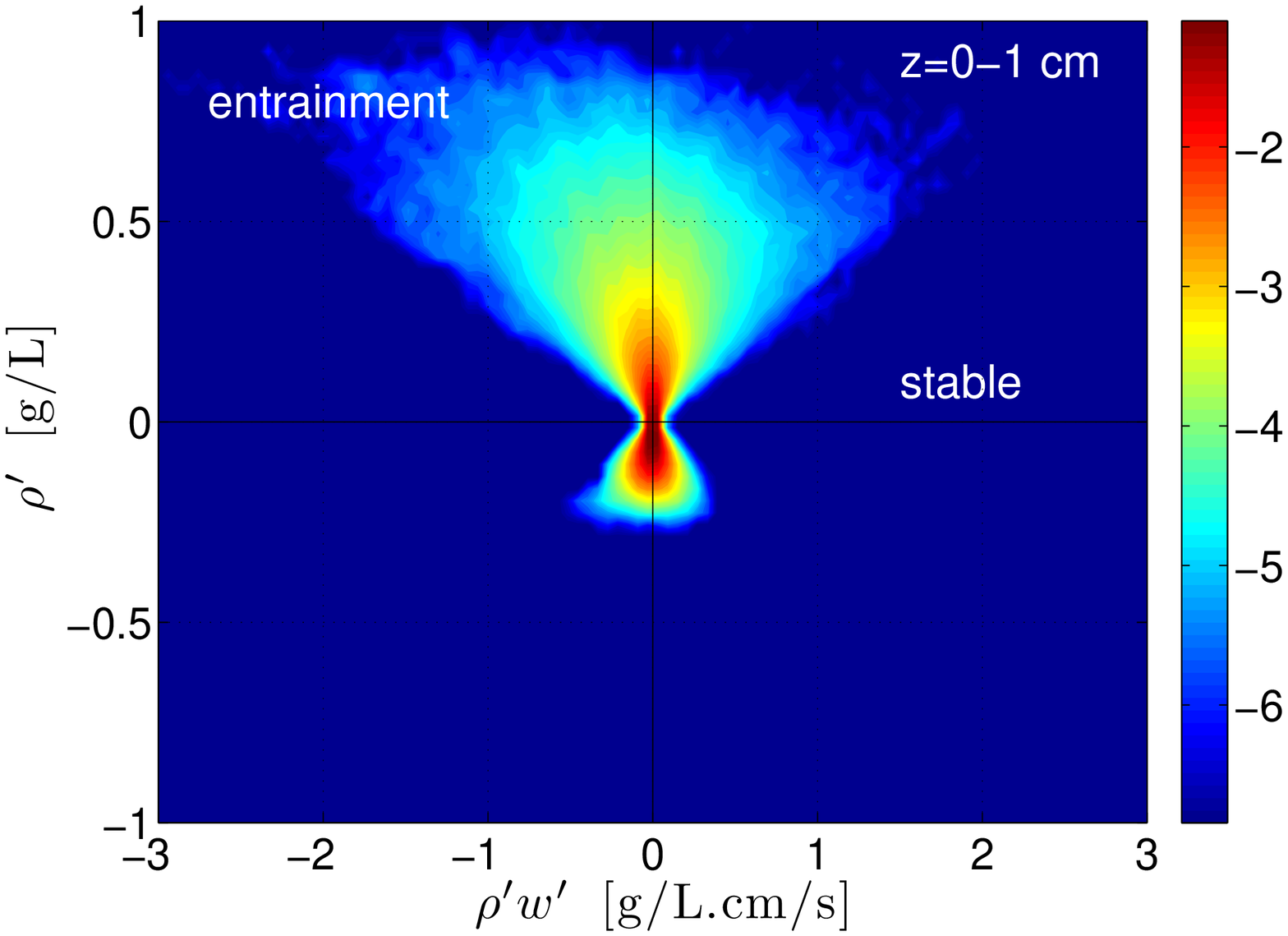}\\
\includegraphics[width=8cm]{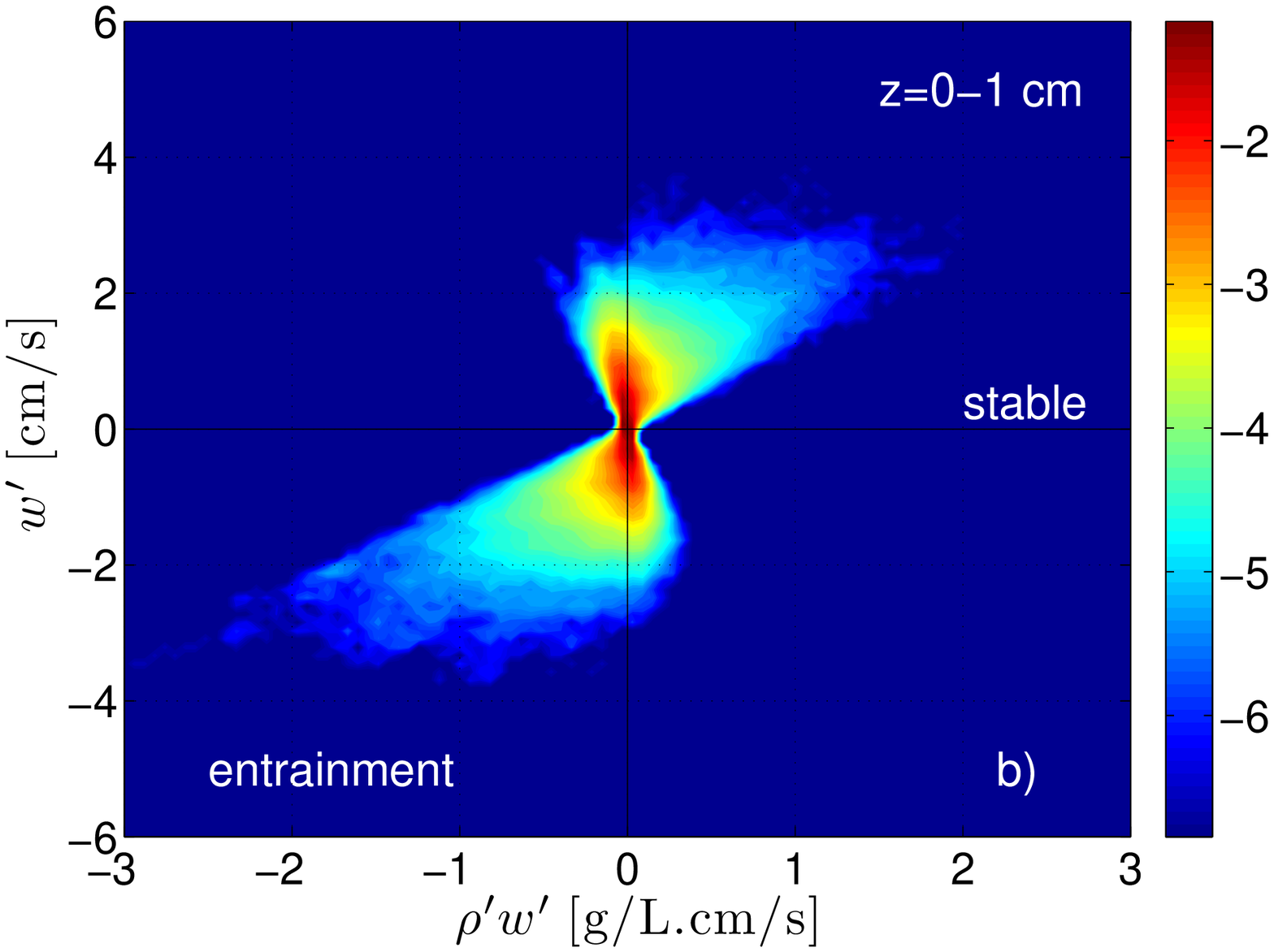}
\caption{\label{fig:2Dhist-2}  2D PDFs of a) density flux vs
density fluctuations and b) density flux vs vertical velocity
fluctuations. Each PDF is constructed using data in a horizontal
band of 1 cm height next to the inclined plate, in a region between 20 and 49 cm from the injection nozzle.}
\end{figure}

\begin{figure}
\centering
\includegraphics[width=9cm]{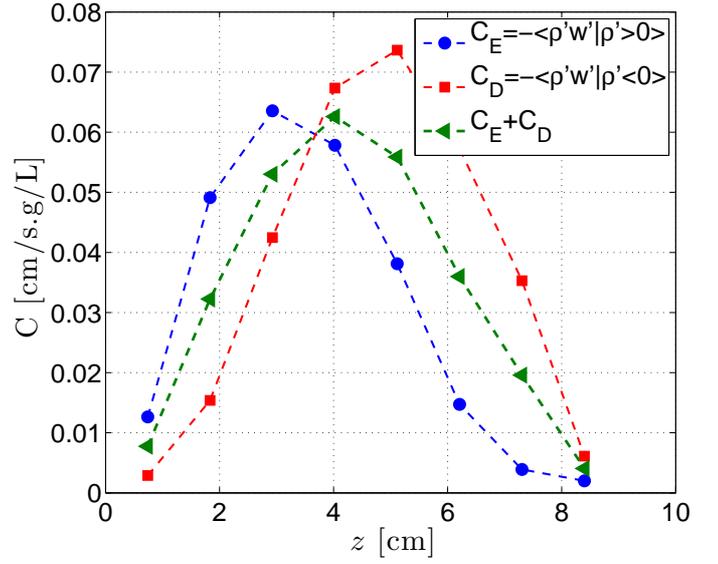}
\caption{\label{fig:ED}  Entrainment and detrainment parameters $C_E$
(circles), $C_D$ (squares), and their average value $(C_E + C_D)/2$ (triangles) as
functions of $z$.}
\end{figure}

The correlations presented in Fig.\ \ref{fig:2Dhist} should also be
reflected in correlations between vertical velocity fluctuations
and density flux.  In Fig.\ \ref{fig:2Dhist-2}, we show a comparison
between $\rho'w' | \rho'$ correlations in Fig.\ \ref{fig:2Dhist}a and
$\rho'w' | w'$ correlations for conditions near the plate, $z = 0-1$ cm. One
gets a similar skewness indicating entrainment (the quadrants for
entrainment and detrainment are flipped vertically for $w'$ compared to $\rho'$)
although there are long tails in the stable region of the $\rho'w' | w'$
correlations. The precise details of the shapes of these PDFs is
beyond the scope of this paper but overall they point to an interesting
approach at a local measure of entrainment and detrainment.  In particular, one
can compute, based on the analysis above, an effective entrainment parameter $C_E = -\langle \rho'w' |
\rho'>0\rangle$ and a corresponding detrainment parameter $C_D = -\langle
\rho'w' | \rho'<0\rangle$ (minus sign added because the fluxes are negative).  In Fig.\
\ref{fig:ED}, we show the quantities $C_E$ and $C_D$ as functions of $z$. 
The results show the behavior described above, namely that entrainment is
dominant near the plate whereas detrainment is larger in the quiescent
fluid far from the plate.  The combined effect of entrainment and
detrainment is described by the addition of the two parameters, i.e,. $C_E
+ C_D$, shown in Fig.\ \ref{fig:ED} as $(C_E + C_D)/2$.  The maximum
density flux happens in the middle of the mixing zone, between 3 and 5
cm. 

\section{Conclusion}
\label{sec:conclu}

In order to allow a comparison of our results with real oceanic situations, some additional features would need to be taken into account. To start with, as in any experimental study of an oceanic process, the Reynolds number dependence needs to be investigated in more detail, since this number in the ocean is far greater than can ever be achieved experimentally. In addition, although the scales at which the oceanic mixing takes place are much smaller than the Rossby deformation radius, it has been shown~\cite{Darelius:DSR:08} that the combined effect of the Coriolis force and the presence of topography (ridges or canyons on the bottom surface) can lead to secondary flows that can contribute to the mixing. Even without the Coriolis force, secondary flows can be created by topographic effects, for example in a curved channel. Finally, the roughness of the bottom (or of the inclined plate in our case) has been shown to influence the behavior of gravity currents~\cite{Baines:JFM:05}. Extensions of this study, taking some of these aspects into account, would contribute to possible extrapolations to oceanic situations. Certain conclusions can already be drawn, however, regarding mixing processes in gravity currents.

Our experimental device provides us with the ability to derive correlations between components of velocity, as well as between velocity and density, thanks to simultaneous measurements of these quantities. In addition, good spatial resolution allows us to study the evolution of these fluxes along both the flow direction and the vertical direction perpendicular to the plate. We measure eddy viscosity and diffusivity, although the constant eddy diffusivity assumption does not work well for the data. A mixing length model provides a better fit of the data with the definition of a typical length associated with the mixing phenomena. Contrary to the eddy diffusivity, this mixing length is very constant in space, allowing us to study its scaling with a length scale constructed from the turbulent dissipation rate $\varepsilon$ and the mean shear, $L_s = (\ol\varepsilon/\la\ol{\dd_z u\ra}^3)^{1/2}$. Using data taken in different configurations of turbulence level and stratification, we show that this scaling is robust, even in the case of a simple wall jet (no stratification), where the mixing is much stronger with a mixing length almost ten times larger than in the stratified cases.

Finally, mixing events can also be observed by looking at the asymmetry of the probability density functions of the momentum and density fluxes. As expected, the strongest mixing takes place at the interface between the current and the ambient fluid. In addition, 2D PDFs of the correlation between the density flux and the density fluctuations provide a local measure of entrainment and detrainment. In particular, we demonstrate that close to the plate the mixing is predominately entrainment of heavier fluid, whereas away from the plate, the mixing is largely associated with detrainment, where the current releases some of its fluid into the ambient medium. Quantitative determinations of entrainment and detrainment are derived from the asymmetry of the PDFs and provide a possible means for obtaining a global measure of net gravity current entrainment from local measurements. Connecting these results with more traditional
measures of entrainment, i.e., mean vertical velocity divided by mean
downstream velocity would be an very interesting extension to
the present work.

We acknowledge useful discussions with H. Aluie, J.-F. Pinton and M. Rivera.  Work performed at Los Alamos National Laboratory was funded by the US Department of Energy under Contract No. DE-AC52-06NA25396.\\





\bibliographystyle{model1-num-names}
\bibliography{mixing.bib}







\end{document}